\documentclass[12pt]{article}

\usepackage{graphicx}
\usepackage{geometry}
\usepackage{tabularx}

\renewcommand{\baselinestretch}{1.3}

\usepackage{amssymb}
\usepackage{amsmath,bm}

\newcommand{\bey}{\begin{eqnarray}}
\newcommand{\eey}{\end{eqnarray}}
\newcommand{\beq}{\begin{equation}}
\newcommand{\eeq}{\end{equation}}

\newcommand{\sgn}{\text{sgn}}
\newcommand{\boldm}[1] {\mathversion{bold}#1\mathversion{normal}}

\newtheorem{theorem}{Theorem}[section]
\newtheorem{lemma}{Lemma}[section]

\begin{document}
\newcolumntype{L}[1]{>{\raggedright\arraybackslash}p{#1}}
\newcolumntype{C}[1]{>{\centering\arraybackslash}p{#1}}
\newcolumntype{R}[1]{>{\raggedleft\arraybackslash}p{#1}}

\begin{center}

\noindent {\large \bf Group least squares regression for linear models with strongly correlated predictor variables}\\[0.2in]
%\noindent {\large \bf A group-based approach for handling multicollinearity from strongly correlated predictor variables in linear models}\\[0.2in]
%\noindent {\large \bf Working with multicollinearity in linear models\\ through a group focused approach}\\[0.2in]
Min Tsao\\
{\footnotesize \em University of Victoria, Canada}
\end{center}

\bigskip

{%\small

\noindent {\bf Abstract:} Traditionally, the least squares regression is mainly concerned with studying the effects of individual predictor variables, but strongly correlated variables generate multicollinearity which makes it difficult to study their effects. Existing methods for handling multicollinearity such as ridge regression are complicated. To resolve the multicollinearity issue without abandoning the simple least squares regression, for situations where predictor variables are in groups with strong within-group correlations but weak between-group correlations, we propose to study the effects of the groups with a group approach to the least squares regression. Using an all positive correlations arrangement of the strongly correlated variables, we first characterize group effects that are meaningful and can be accurately estimated. We then present the group approach with numerical examples and demonstrate its advantages over existing methods for handling multicollinearity. We also address a common misconception about prediction accuracy of the least squares estimated model and discuss through an example similar group effects in generalized linear models.

\bigskip
\bigskip

}

%\footnotetext{Min Tsao is Professor, Department of Mathematics and Statistics, University of Victoria, British Columbia, Canada V8W 3R4 (Email: mtsao@uvic.ca).}

\section{Introduction}
\label{intro}

Multicollinearity due to strongly correlated predictor variables is a long-standing problem without a satisfactory solution. It arises frequently in observational studies in social sciences and medical research. In this paper, we show that multicollinearity per se is not a problem; the problem is that what we have been trying to do with the strongly correlated variables are misguided and unattainable. We also present a solution based on appropriate use of such variables. To introduce the problem, consider multiple regression model
\begin{equation}
\mathbf{y}=\mathbf{X}\bm{\beta} +\bm{\varepsilon}, \label{m0}
\end{equation}
where $\mathbf{y}$ is an $n$-vector of observations, $\mathbf{X}=[\mathbf{1}_n,\mathbf{x}_1,\dots,\mathbf{x}_{p}]$ is a
known $n\times (p+1)$ design matrix with $p\geq 2$ and $\mathbf{1}_n$ being the $n$-vector of 1's, $\boldsymbol{\beta}=(\beta_0,\beta_1,\dots,\beta_{p})^T$ is the unknown vector of regression parameters, and $\boldsymbol{\varepsilon}$ is an $n$-vector of i.i.d. normal random errors with mean $0$ and variance $\sigma^2$. Throughout this paper, we work under the low dimensional setting where $n>p$ and $rank(\mathbf{X})=p+1$ so that the least squares estimator for $\boldsymbol{\beta}$,
\begin{equation}
\hat{\boldsymbol{\beta}}=(\hat{\beta}_0,\hat{\beta}_1,\dots,\hat{\beta}_{p})^T= (\mathbf{X}^T\mathbf{X})^{-1}\mathbf{X}^T\mathbf{y}, \label{ole}
\end{equation}
is available. We assume that the $p$ predictor variables can be partitioned into $k$ groups $\{\mathbf{X}_i\}^k_{i=1}$ such that ($i$) there is at least one group with 2 or more variables, ($ii$) variables in the same group are strongly correlated, and ($iii$) variables from different groups are weakly correlated. Let $\boldsymbol{\beta}_i$ be the parameter vector for variables in group $\mathbf{X}_i$. Model (\ref{m0}) may be written as 
\begin{equation}
\mathbf{y}=\beta_0\mathbf{1}_n+\mathbf{X}_1\boldsymbol{\beta}_1+\mathbf{X}_2\boldsymbol{\beta}_2 +\dots+\mathbf{X}_k\boldsymbol{\beta}_k + \boldsymbol{\varepsilon}. \nonumber \label{m1}
\end{equation}
Here, $\boldsymbol{\beta}_i$ reduces to a scalar if there is only 1 variable in group $\mathbf{X}_i$. 
Let $\hat{\boldsymbol{\beta}}_i$ be the least squares estimator for $\boldsymbol{\beta}_i$.
When there are 2 or more variables in $\mathbf{X}_i$, their strong correlations generate multicollinearity which makes variances of elements of $\hat{\boldsymbol{\beta}}_i$ large, rendering $\hat{\boldsymbol{\beta}}_i$ a poor estimator for ${\boldsymbol{\beta}}_i$. 

There is a large body of literature on detecting and handling the multicollinearity problem; see, for example, Draper and Smith (1998), Belsley, Kuh and Welsch (2004), Montgomery, Peck and Vining (2012). Here,  we only briefly discuss the main methods for handling the problem.
The most well-known methods are the ridge regression (Hoerl and Kennard, 1970) and principal component regression (Jolliffe, 1986). There are also other methods such as latent root regression (Webster, Gunst and Mason, 1974) and model respecification by eliminating some predictor variables. There have been a number of studies that evaluate these methods including Hoerl, Kennard and Boldwin (1975), Lawless and Wang (1976), Gunst, Webster and Mason (1976), Dampster, Schatzoff and Wermuth (1977), Gunst and Mason (1977) and Lawless (1978).  One of the main criteria used for evaluation is the mean squared error of an estimator $\tilde{\boldsymbol{\beta}}$ for $\boldsymbol{\beta}$, $E[(\tilde{\boldsymbol{\beta}}-\boldsymbol{\beta})^T(\tilde{\boldsymbol{\beta}}-\boldsymbol{\beta})]$.
Estimators given by these methods are biased, but they are capable of achieving smaller mean squared error than the least squares estimator $\hat{\boldsymbol{\beta}}$.
However, except for this advantage, these estimators are difficult to use because their sampling properties are in general not available as they depend on the data in complicated ways. The ridge regression estimator, for example, involves a penalty parameter whose value is usually determined by cross-validation. The distribution of the penalty parameter and thus that of the estimator are unavailable. It is also difficult to choose among these methods as extensive comparisons have found no single best overall method;  see Montgomery, Peck and Vining (2012) for more discussion. Further, some authors such as Conniffe and Stone (1973) are critical of biased estimation methods. Draper and Van Nostrand (1979) identified two cases where ridge regression may be appropriate but also recommended against the use of biased estimation methods in general. Nevertheless, these methods are still the most used tools for handling multicollinearity.

Is multicollinearity really such an insurmountable problem for the least squares regression that we have to abandon this simple method of regression in favour of complicated alternatives? Traditionally, the focus of regression analyses has been on the impact of individual predictor variables. For example, in estimation, the focus has been on estimating parameters of individual variables; in variable selection, it has been on inclusion or exclusion of individual variables. With this focus on individual variables, multicollinearity has been a problem for the least squares regression as it cannot accurately estimate parameters of the strongly correlated variables which in turn leads to difficulties in variable selection and prediction. Nevertheless, we argue that neither multicollinearity nor the least squares regression is responsible for these problems; the wrong focus on the impact of individual variables is the real culprit. In Remark [a] of Section 2.3, we note that estimating the parameter of a variable in a strongly correlated group is a form of extreme extrapolation. That it cannot be done accurately is solely the consequence of extrapolating far beyond the data range.
Strongly correlated variables appear naturally in groups. Individual parameters of these variables are not meaningful.  Instead of focusing on their individual impact, we should respect their group nature by handling them in groups and focusing on their collective impact on the response variable. To this end, we propose a group approach to the least squares regression which still relies on $\hat{\boldsymbol{\beta}}$ but differs from the traditional least squares regression in three aspects: ($i$) for a group $\mathbf{X}_i$ with 2 or more variables, the group approach will not attempt to estimate or make inference about individual elements of $\boldsymbol{\beta}_i$; instead, it will focus on estimation and inference for those linear combinations of the elements of $\boldsymbol{\beta}_i$ that represent meaningful group effects of $\mathbf{X}_i$; ($ii$) it will perform variable selection at the group level in that variables in a group  $\mathbf{X}_i$ are either all in or all out; and ($iii$) it will analyse prediction accuracy of the least squares estimated model through group effects. For a group $\mathbf{X}_i$ with only 1 variable, its group effect is the parameter of the variable, so the group approach will still estimate and make inference of the parameter just like in the traditional least square regression.

Comparing to existing methods for handling multicollinearity, the group approach to the least squares regression has the advantage that it is very simple in computation and its theories for estimation, inference and prediction are already in place as it is still least squares regression with only a change of focus from individual to group effects for strongly correlated variables. In contrast, computation for the ridge regression and principal component regression are more complicated and theories for these methods are convoluted and even intractable. Additional advantages of the group approach include ($i$) it retains the simple least squares estimators $\hat{\beta}_i$; those for variables not strongly correlated with others are good unbiased point estimators of their parameters we can still use; those for strongly correlated variables are only used for estimation and inference of group effects of such variables and making predictions, but they are not used as point estimators as parameters of such variables are not estimated under the group approach; ($ii$) the regression mean squared error remains a good unbiased estimator for the error variance $\sigma^2$, and ($iii$) existing (non-group based) methods of inference, variable selection and model diagnosis for the least squares regression may be adopted with a minor adjustment of handling strongly correlated variables in groups. The ridge regression and principal component regression have none of these advantages.

There is a widely held view that when there is multicollinearity in the data, alternative regression methods in general and the ridge regression in particular give more accurate predictions than the least squares regression. Although there is no proof to support this view, it has appeared in many papers, books and internet sites. Through a group effect based analysis on the prediction accuracy of the least squares estimated model and a comparison with the ridge regression, we show that this is a misconception arising from comparing prediction accuracy at points where predictions are not meaningful and should not be made. At points where predictions are meaningful, the least squares regression is actually more accurate than the ridge regression. 

The rest of this paper is organized as follows. In Section 2, we discuss group effects of strongly correlated variables and characterize group effects that can be accurately estimated. We also discuss why such group effects are meaningful but individual parameters of these variables are not. In Section 3, we present the group approach through numerical examples and discuss estimation, inference, variable selection and prediction under this approach. We also provide ($i$) a comparison of group versus non-group based variable selection, ($ii$) a comparison on prediction accuracy between the least squares regression and the ridge regression and ($iii$) a new characterization for the region in the predictor variable space over which predictions by the least squares estimated model are accurate. The misconception is discussed near the end of Section 3.3. In Section 4, we apply the group approach to Hald cement data to illustrate several points. We conclude with a few remarks in Section 5. Proofs of lemmas and theorems are in Appendix I. In Appendix II, we give an example of the impact of multicollinearity on generalized linear models.

\section{Group effects of strongly correlated predictor variables} %22222222222222222222222222222222222222222222222222222222222222222222222222222222222222222222222222222222

Group effects lie at the heart of the group approach to the least squares regression. Tsao (2019) studied estimation of group effects in a theoretical model containing strongly correlated predictor variables with a restrictive uniform correlation structure. We now revisit the estimation problem without imposing any parametric correlation structure on the strongly correlated variables and generalize results in Tsao (2019) to all linear models. 
For this section, we let $\mathbf{X}_1=[\mathbf{x}_1,\mathbf{x}_2,\dots,\mathbf{x}_q]$ and $\mathbf{X}_2=[\mathbf{x}_{q+1},\mathbf{x}_{q+2},\dots,\mathbf{x}_{p}]$, and write (\ref{m0}) as
\begin{equation}
\mathbf{y}=\beta_0\mathbf{1}_n+\mathbf{X}_1\boldsymbol{\beta}_1+\mathbf{X}_2\boldsymbol{\beta}_2 +\boldsymbol{\varepsilon}, \label{m2}
\end{equation}
where $2\leq q\leq p$, $\boldsymbol{\beta}_1=(\beta_1,\beta_2,\dots,\beta_q)^T$, $\boldsymbol{\beta}_2=(\beta_{q+1},\beta_{q+2},\dots,\beta_{p})^T$,
and $\mathbf{X}_1$ is a group of strongly correlated variables satisfying
($i$) for $1\leq i, j \leq q$, absolute values of $r_{ij}=corr(\mathbf{x}_i,\mathbf{x}_j)$ are all above $\frac{\sqrt{2}}{2}$  ($\approx 0.71$) and ($ii$) variables in $\mathbf{X}_1$ are not strongly correlated with variables in $\mathbf{X}_2$. Condition ($i$) is needed to ensure that variables in $\mathbf{X}_1$ will all have positive correlations after appropriate sign changes; see equation (\ref{apc}). For this section, $\mathbf{X}_2$ holds all variables not in $\mathbf{X}_1$. There may be more strongly correlated groups among variables in $\mathbf{X}_2$ but it suffices to study the group effects of just $\mathbf{X}_1$ as results obtained apply to all such groups.
Consider the class of linear combinations of $\beta_1, \beta_2, \dots, \beta_q$,
\beq
\Xi=\{\xi(\mathbf{w})\hspace{0.05in}|\hspace{0.05in}\xi(\mathbf{w})=w_1\beta_1+w_2\beta_2+\dots+w_q\beta_q \}, \label{g.eff}
\eeq
where $\mathbf{w}=(w_1,w_2,\dots,w_q)^T$ is any $q$-vector satisfying $\sum_{i=1}^q |w_i|=1$. Set $\Xi$ is the class of normalized group effects of variables in $\mathbf{X}_1$. Each $\xi(\mathbf{w})$ in $\Xi$ is a (normalized) group effect and the corresponding vector $\mathbf{w}$ is its weight vector. An effect $\xi(\mathbf{w})$ has the interpretation as the expected change in the response variable ${y}$ when the $q$ predictor variables in $\mathbf{X}_1$ change by the amount $\mathbf{w}$; that is, $x_1, x_2,\dots,x_q$ change by the amount $w_1, w_2, \dots, w_q$, respectively, at the same time. In this sense, we say that $\xi(\mathbf{w})$ represents a collective impact or a group effect on $y$. Not all group effects can be accurately estimated and some group effects are not meaningful. For example, $\beta_1$ is a special group effect with $w_1=1$ and $w_j=0$ for $j\neq 1$, but it cannot be accurately estimated. It is also not a meaningful effect (see Remark [a]). We now characterize group effects that can be accurately estimated.
To this end, we first introduce an all positive correlations arrangement of the strongly correlated variables
% develop a simple way to quantify the level of multicollinearity caused by a group of strongly correlated variables 
and then study the limiting properties of their correlation matrix.

\subsection{All positive correlations arrangement of strongly correlated variables and limiting properties of their correlation matrix}

Let $\mathbf{R}$ be the full rank correlation matrix of $\mathbf{x}_1, \mathbf{x}_2, \dots, \mathbf{x}_q$,
\beq
\mathbf{R} =
\left[ \begin{array}{cccc}
1 & r_{12}  & \cdots & r_{1q} \\
r_{21} & 1 & \cdots & r_{2q}  \\
\cdot&\cdot&\cdots &\cdot \\
r_{q1} & r_{q2} & \cdots & 1 \\
\end{array} \right]_{q\times q}.  \label{corr.m}
\eeq
Some of the $r_{ij}$ may be negative but since all $|r_{ij}|$ are above $\frac{\sqrt{2}}{2}$, let $\sgn(r_{1j})$ be the sign of $r_{1j}=corr(\mathbf{x}_1,\mathbf{x}_j)$ for $j=2,3,\dots, q$, by Theorem 3.1 in Tsao (2019) the following signed version of the set of $q$ variables
\begin{equation}
\mathbf{x}_1,\sgn(r_{12})\mathbf{x}_2,\dots,\sgn(r_{1q})\mathbf{x}_q  \label{apc}
\end{equation}
satisfies that all pairwise correlations are positive. We call (\ref{apc}) an {all positive correlations (APC) arrangement} of $\mathbf{x}_1, \mathbf{x}_2, \dots, \mathbf{x}_q$. For the rest of this section, we assume that these $q$ variables are already in an APC arrangement so that all $r_{ij}$ in (\ref{corr.m}) are positive. If they are not in an APC arrangement, we can replace them with their APC version (\ref{apc}); see Section 4 for an example.

The importance of using the APC arrangement is twofold. Firstly, it makes it convenient to identify important and meaningful group effects in Sectoin 2.3. Secondly, it makes it easy to measure the level of multicollinearity generated by the $q$ variables and to formulate the question of interest. To see the second point, let $r_M=\min\{r_{ij}\}$. Under the APC arrangement, all $r_{ij}$ satisfy $0<r_M\leq r_{ij}< 1$, so when $r_M$ goes to 1, all $r_{ij}$ go to 1 which makes the multicollinearity stronger. In this sense, an increase in $r_M$ represents an increase in the level of multicollinearity, so we will use $r_M$ to measure this level. Our question of interest can now be formulated as that of identifying group effects in (\ref{g.eff}) that can be accurately estimated when $r_M$ is close to 1. 

To answer the above question, we first study the limiting properties of $\mathbf{R}$ and $\mathbf{R}^{-1}$ when $r_M $ approaches 1.
Since $\mathbf{R}$ is a correlation matrix, it is positive definite, so it has $q$ positive eigenvalues $\lambda_1\geq \lambda_2 \geq \dots\geq \lambda_q>0$. Let $\mathbf{v}_1, \mathbf{v}_2,\dots,\mathbf{v}_q$ be their corresponding orthonormal eigenvectors, respectively, and $\mathbf{1}_q$ be the $q$-vector of 1's. We have the following results.

\begin{lemma} \label{lemma1}
Correlation matrix $\mathbf{R}$ satisfies
\begin{itemize}
\item[(i)] $\lambda_1 \rightarrow q$ and $\lambda_i \rightarrow 0$ for $i=2,3,\dots,q$  as $r_M \rightarrow 1$; and
\item[(ii)] $\mathbf{v}_1 \rightarrow \frac{1}{\sqrt{q}} \mathbf{1}_q$ as $r_M \rightarrow 1$.
\end{itemize}
\end{lemma}

\begin{lemma} \label{lemma2}
The inverse matrix $\mathbf{R}^{-1}$ satisfies
\begin{itemize}
\item[(i)] $\mathbf{v}_1^T\mathbf{R}^{-1} \mathbf{v}_1 > \frac{1}{q}$; and
\item[(ii)] $\mathbf{v}_1^T\mathbf{R}^{-1} \mathbf{v}_1 \rightarrow  \frac{1}{q}$ as $r_M \rightarrow 1$.
\end{itemize}
\end{lemma}

The proofs of these lemmas are in the Appendix.

\subsection{The eigen-effect of strongly correlated predictor variables} %----------------------------------------------------------------

In this section, we identify one group effect for the standardized version of (\ref{m2})  that can be very accurately estimated at high levels of multicollinearity. It will be used to identify other effects that can be accurately estimated.

Let $\mathbf{x}_i=(x_{1i},x_{2i},\dots,x_{ni})^T$, $\bar{x}_i=\frac{1}{n}\sum^n_{j=1}x_{ji}$ and $s_i^2=\sum^n_{j=1}(x_{ji}-\bar{x}_i)^2$ which is $(n-1)$ times the sample variance of $\mathbf{x}_i$. We call
\beq
\mathbf{x}_i'=\frac{\mathbf{x}_i-\bar{x}_i\mathbf{1}_n}{s_i}  \label{std.x}
\eeq
the standardized variable which has mean zero and length one. Let $\mathbf{y}=(y_{1},y_{2},\dots,y_{n})^T$, $\bar{y}=\frac{1}{n}\sum^n_{j=1}y_{j}$ and $\mathbf{y}'=\mathbf{y}-\bar{y}$. We can write (\ref{m2}) as
\begin{equation}
\mathbf{y}'=\mathbf{X}'_1\boldsymbol{\beta}'_1+\mathbf{X}'_2\boldsymbol{\beta}'_2 +\boldsymbol{\varepsilon}, \label{m3}
\end{equation}
where $\mathbf{X}_1'=[\mathbf{x}_1',\mathbf{x}'_2,\dots,\mathbf{x}'_q]$, $\mathbf{X}'_2=[\mathbf{x}'_{q+1},\mathbf{x}_{q+2}',\dots,\mathbf{x}'_{p}]$, $\boldsymbol{\beta}'_1=(\beta'_1,\beta'_2,\dots,\beta'_q)^T$, and $\boldsymbol{\beta}'_2=(\beta'_{q+1},\beta'_{q+2},\dots,\beta'_{p})^T$. We call model (\ref{m3}) the {standardized model}. The relationship between parameters in models (\ref{m3}) and (\ref{m2}) is
\beq
{\beta}_0=\bar{y}-\sum^{p}_{i=1}\bar{x}_i{\beta}_i'/s_i \mbox{\hspace{0.1in} and \hspace{0.1in}} {\beta}_i={\beta}_i'/s_i
\mbox{\hspace{0.1in} for $i=1,2,\dots,p$}. \label{coeffs.1}
\eeq
Let $\mathbf{X}'=[\mathbf{X}_1',\mathbf{X}_2']$. Then, $\mathbf{X}'^T\mathbf{X}'=[r_{ij}]\in \mathbb{R}^{p\times p}$ is the correlation matrix of the  $p$ predictor variables in models (\ref{m3}) or (\ref{m2}) where $r_{ij}=corr(\mathbf{x}'_i,\mathbf{x}'_j)=corr(\mathbf{x}_i,\mathbf{x}_j)$. Partition this correlation matrix as follows:
\beq
\mathbf{X}'^T\mathbf{X}'=
\left[ \begin{array}{cc}
\mathbf{R}_{11} &\mathbf{R}_{12} \\
\mathbf{R}_{21} & \mathbf{R}_{22}  \\
\end{array} \right]_{p\times p},  \label{corr.total} 
\eeq
where $\mathbf{R}_{11}=\mathbf{R} \in \mathbb{R}^{q\times q}$ is the correlation matrix (\ref{corr.m}) of the $q$ variables in $\mathbf{X}_1'$, and $\mathbf{R}_{12}$ is the between-group correlation matrix of $\mathbf{X}_1'$ and $\mathbf{X}_2'$. By (\ref{corr.total}),
{\small
\beq
[\mathbf{X}'^T\mathbf{X}']^{-1}=
\left[ \begin{array}{cc}
[\mathbf{R}_{11}-\mathbf{R}_{12}\mathbf{R}^{-1}_{22}\mathbf{R}_{21}]^{-1} &  \mathbf{R}_{11}^{-1}\mathbf{R}_{12}[\mathbf{R}_{21}\mathbf{R}^{-1}_{11}\mathbf{R}_{12} - \mathbf{R}_{22}]^{-1}  \\

[\mathbf{R}_{21}\mathbf{R}^{-1}_{11}\mathbf{R}_{12} - \mathbf{R}_{22}]^{-1} \mathbf{R}_{21} \mathbf{R}_{11}^{-1} & [\mathbf{R}_{22}-\mathbf{R}_{21}\mathbf{R}^{-1}_{11}\mathbf{R}_{12}]^{-1} \\
\end{array} \right].  \label{corr.inv}
\eeq
}

Let $\mathbf{R}^*=[\mathbf{R}_{11}-\mathbf{R}_{12}\mathbf{R}^{-1}_{22}\mathbf{R}_{21}]$. Then, $\mathbf{R}^*$ is a symmetric positive definite matrix as ${\mathbf{R}^*}^{-1}$ is a diagonal block of the positive definite matrix $[\mathbf{X}'^T\mathbf{X}']^{-1}$ in (\ref{corr.inv}). Let $\lambda^*_{1}$ be its largest eigenvalue and $\mathbf{v}_1^*=(v_{11}^*, v_{12}^*, \dots, v_{1q}^*)^T$ be the corresponding orthonormal eigenvector. 
We call linear combination
\begin{equation}
\xi_E={\mathbf{v}^*_1}^T\boldsymbol{\beta}'_1=v_{11}^*\beta_1'+v_{12}^*\beta_2'+\dots+v_{1q}^*\beta_q'  \label{e.eff}
\end{equation}
the eigen-effect. Since $\|\mathbf{v}^*_1\|=1$, $1\leq  \sum^q_{i=1} |v_{1i}^*| \leq \sqrt{q}$ and so $\xi_E$ may not be a normalized effect.  Nevertheless, for technical convenience we will first study $\xi_E$ and will give a simple normalized representation of $\xi_E$ later.

Let $\hat{\boldsymbol{\beta}}'=(\hat{\beta}'_1, \hat{\beta}'_2,\dots, \hat{\beta}'_p)^T$ be the least squares estimator for ${\boldsymbol{\beta}}'=({\boldsymbol{\beta}'_1}^T,{\boldsymbol{\beta}_2'}^T)^T$. The minimum-variance unbiased linear estimator for $\xi_E$ is
\beq
\hat{\xi}_E={\mathbf{v}^*_1}^T\hat{\boldsymbol{\beta}}'_1=v_{11}^*\hat{\beta}'_1+v_{12}^*\hat{\beta}'_2+\dots+v_{1q}^*\hat{\beta}'_q.
\label{e.eff.est}
\eeq
Since $\hat{\xi}_E$ is an unbiased estimator for ${\xi}_E$, it is accurate if $var(\hat{\xi}_E)$ is small.
Although none of the $\beta'_i$ in (\ref{e.eff}) is accurately estimated by $\hat{\beta}'_i$ in (\ref{e.eff.est}) when $r_M$ is high, the following theorem shows ${\xi}_E$ is accurately estimated by $\hat{\xi}_E$.

%%%%%%%%%%%%%%%%%%%%%%%%%%%%%%%%%%%%%%%%%%%%%%%%%%% Theorem 111111111111111111111111111
\begin{theorem} \label{thm1}
For the group of strongly correlated variables in $\mathbf{X}_1'$ in (\ref{m3}),
\begin{itemize}
\item[(i)] if they are uncorrelated with variables in $\mathbf{X}_2'$, then ($i_1$) $var(\hat{\xi}_E) > \sigma^2/q$ and  ($i_2$) $var(\hat{\xi}_E) \rightarrow \sigma^2/q$ as $r_M\rightarrow 1$; and \vspace{0.05in}
\item[(ii)] if they are correlated with variables in $\mathbf{X}_2'$ but the between-group correlation matrix $\mathbf{R}_{12}\rightarrow \mathbf{0}$
 as $r_M\rightarrow 1$, then $var(\hat{\xi}_E) \rightarrow \sigma^2/q$ as $r_M\rightarrow 1$. \footnote{$\mathbf{R}_{12}\rightarrow \mathbf{0}$ denotes element-wise convergence of $\mathbf{R}_{12}$ to zero. It implies $\mathbf{R}_{12}\mathbf{R}^{-1}_{22}\mathbf{R}_{21} \rightarrow \mathbf{0}$ under general conditions such as $\|\mathbf{R}^{-1}_{22}\|_{max}$ is bounded or $(\|\mathbf{R}_{12}\|_{max})^2(\|\mathbf{R}^{-1}_{22}\|_{max})=o(1)$. This observation will be used in the proof of ($ii$) which requires $\mathbf{R}_{12}\mathbf{R}^{-1}_{22}\mathbf{R}_{21} \rightarrow \mathbf{0}$. }
%Section 3.2 contains a real data example illustrating $\mathbf{R}_{12}\rightarrow \mathbf{0}$ implies $\mathbf{R}_{12}\mathbf{R}^{-1}_{22}\mathbf{R}_{21 \rightarrow \mathbf{0}$.}
\end{itemize}
\end{theorem}
%%%%%%%%%%%%%%%%%%%%%%%%%%%%%%%%%%%%%%%%%%%%%%%%%%
To interpret Theorem \ref{thm1}, when variables in  $\mathbf{X}_1'$ are uncorrelated with those in  $\mathbf{X}_2'$, result ($i_1$) gives a lower bound on $var(\hat{\xi}_E)$ and result ($i_2$) shows $var(\hat{\xi}_E)$ approaches this lower bound as $r_M$ approaches its upper bound $1$. Thus, $\xi_E$ is more accurately estimated by $\hat{\xi}_E$ at higher levels of multicollinearity. 
Result ($ii$) gives the asymptotic behaviour of $var(\hat{\xi}_E)$ when $r_M$ goes to 1 and correlations between variables in $\mathbf{X}_1'$ and $\mathbf{X}_2'$ go to zero ($\mathbf{R}_{12}\rightarrow \mathbf{0}$). It implies that when such correlations are weak and the level of multicollinearity is high, $var(\hat{\xi}_E)$ is approximately $\sigma^2/q$. 
%The proof of Theorem \ref{thm1} is in the Appendix.

Theorem \ref{thm1} does not cover the case where some variables in $\mathbf{X}_1'$ are strongly correlated with some variables in $\mathbf{X}'_2$.
We are not interested in this case as it weakens the notion of $\mathbf{X}_1'$ being a (stand-alone) group of strongly correlated variables which renders its group effects not meaningful. Turning now to other effects defined by unit vectors that may be accurately estimated when $r_M$ is high, the following result shows where such effects may be found.
%%%%%%%%%%%%%%%%%%% Theorem 222222222222222222222222222222222222222222222222222222222222222222222222222
\begin{theorem} \label{thm2}
For $\delta>0$, define a neighbourhood of $\mathbf{v}^*_1$ on the unit sphere
\begin{equation} 
{\mathcal N}_{\delta}= \{\mathbf{v}\in \mathbb{R}^q: \|\mathbf{v}\|=1 \mbox{\hspace{0.01in} and \hspace{0.01in}} \sqrt{1-\delta}< \mathbf{v} \cdot \mathbf{v}^*_1 \leq 1  \}. \label{temp11}
\end{equation}
Suppose the between-group correlation matrix $\mathbf{R}_{12}\rightarrow \mathbf{0}$ as $r_M\rightarrow 1$.  If a unit vector $\mathbf{v} \notin {\cal N}_\delta$, then $var(\mathbf{v}^T\hat{\boldsymbol{\beta}}'_1) \rightarrow \infty$ as $r_M\rightarrow 1$.
\end{theorem}

\subsection{Characterization of group effects that can be accurately estimated} %----------------------------------------------------------------

Theorem \ref{thm2} implies that all $\mathbf{v}^T\boldsymbol{\beta}'_1$ that can be accurately estimated at high $r_M$ levels are given by $\mathbf{v} \in {\cal N}_\delta$.  
Let $s(\mathbf{v})$ be the sum of absolute values of elements of $\mathbf{v}$. Then, $1\leq s(\mathbf{v}) \leq \sqrt{q}$ and $\mathbf{w}=\mathbf{v}/s(\mathbf{v})$ is a bijection that maps $ {\cal N}_\delta$ into a small open neighbourhood of the normalized eigenvector $\mathbf{v}^*_1/s(\mathbf{v}^*_1)$ on the simplex $\sum_{i=1}^q w_i =1$. Weight $\mathbf{w}$ of group effects that can be accurately estimated are in this open neighbourhood. In this sense, such effects are in a neighbourhood of the normalized eigen-effect $\xi^*_E=\xi_E/s(\mathbf{v}^*_1)$.

To identify a simpler effect to represent $\xi_E^*$ and its neighbourhood, when variables in $\mathbf{X}_1'$ are uncorrelated with variables in $\mathbf{X}_2'$, $\mathbf{R}_{12}=\mathbf{0}$ and $\mathbf{R}^*=\mathbf{R}$, so $\lambda^*_1=\lambda_1$ and $\mathbf{v}_1^*=\mathbf{v}_1$. By Lemma 1,  $\mathbf{v}_1 \rightarrow \frac{1}{\sqrt{q}} \mathbf{1}_q$ as $r_M \rightarrow 1$, which implies $s(\mathbf{v}_1) \rightarrow \sqrt{q}$ and $\mathbf{v}_1/s(\mathbf{v}_1) \rightarrow  \frac{1}{q} \mathbf{1}_q$. When variables in $\mathbf{X}_1'$ and $\mathbf{X}_2'$ are correlated, $\mathbf{v}_1^* \rightarrow \frac{1}{\sqrt{q}} \mathbf{1}_q$ and thus $\mathbf{v}_1^*/s(\mathbf{v}_1^*) \rightarrow  \frac{1}{q} \mathbf{1}_q$ also hold under general conditions (see proof of Theorem \ref{thm1}($ii$)). Thus, $\xi_E^* \rightarrow \xi_A$ as $r_M\rightarrow 1$ where
\beq
\xi_A=\frac{1}{q}\mathbf{1}_q^T\boldsymbol{\beta}'_1= \frac{1}{q} ( \beta_1'+\beta_2'+\cdots+\beta_q'). \label{ave.eff}
\eeq
We call $\xi_A$ the average group effect of the $q$ strongly correlated variables in $\mathbf{X}_1'$. The minimum-variance unbiased linear estimator for ${\xi}_A$ is
\beq
\hat{\xi}_A=\frac{1}{q}\mathbf{1}_q^T\hat{\boldsymbol{\beta}}'_1=\frac{1}{q}(\hat{\beta}'_1+\hat{\beta}'_2+\dots+\hat{\beta}'_q).
\label{eeff.s}
\eeq
When $r_M$ is close to 1, $\hat{\xi}_A\approx \hat{\xi}^*_E$ and so $var(\hat{\xi}_A)\approx var(\hat{\xi}_E^*)=var(\hat{\xi}_E)/[s(\mathbf{v}^*_1)]^2$. Theorem 2.1 and $s(\mathbf{v}_1^*) \rightarrow \sqrt{q}$ then imply that $var(\hat{\xi}_A)\approx \sigma^2/q^2$. On the other hand, when all variables are uncorrelated, $var(\hat{\xi}_A)=\sigma^2/q$.  This shows that the estimation of $\xi_A$ benefits from a high level of multicollinearity in that it makes $var(\hat{\xi}_A)$ approximately $q$ times smaller.
Our subsequent discussions will be centred on $\xi_A$ as it has simpler expression and interpretation than $\xi_E^*$.

For the unstandardised model (\ref{m2}) where $\beta_1, \beta_2, \cdots, \beta_q$ are parameters of the strongly correlated variables in $\mathbf{X}_1$, let $\mathbf{w}^*=(w_1^*,w_2^*,\dots,w_q^*)^T$ where
\beq
w^*_i=\frac{s_i}{\sum^q_{j=1}s_j}  \label{weights}
\eeq
for $i=1,2,\dots,q$. We call the following weighted average
\beq
{\xi}_W=w_1^*{\beta}_1+w_2^*{\beta}_2+\dots+w_p^*{\beta}_q  \label{var.ave}
\eeq
the variability weighted average effect of the variables in $\mathbf{X}_1$ as $w_i^*$ is proportional to the variability (measured by $s_i$) of $\mathbf{x}_i$. Using the least squares estimator in (\ref{ole}),  the minimum-variance unbiased linear estimator for $\xi_W$ is
\beq
\hat{\xi}_W=w_1^*\hat{\beta}_1+w_2^*\hat{\beta}_2+\dots+w_p^*\hat{\beta}_q. \label{est2}
\eeq
Noting that relationship (\ref{coeffs.1}) between the coefficients of the original and standardized models also applies to their respective least squares estimates, $\hat{\xi}_W$ can be expressed in terms of $\hat{\xi}_A$ as
\beq
\hat{\xi}_W=\frac{1}{\sum^q_{j=1}s_j} \sum_{i=1}^q s_i\hat{\beta}_i = \frac{1}{\sum^q_{j=1}s_j}\left(\sum_{i=1}^q \hat{\beta}_i'\right)
=\frac{q}{\sum^q_{j=1}s_j}\hat{\xi}_A.   \label{rel}
\eeq
When $r_M$ is close to 1, since $var(\hat{\xi}_A)$ is approximately $\sigma^2/q^2$, (\ref{rel}) implies
\beq
var(\hat{\xi}_W)=\left(\frac{q}{\sum^q_{j=1}s_j}\right)^2var(\hat{\xi}_A) \approx \frac{\sigma^2}{\left(\sum^q_{i=1}s_i\right)^2}. \nonumber %\label{eff.var}
\eeq
In practice, $(\sum^q_{i=1}s_i)^2$ is usually large, so $var(\hat{\xi}_W)$ is much smaller than $\sigma^2$. Using $\xi_A$ and $\xi_W$ as reference points, we now characterize the set of effects that are meaningful and can be accurately estimated. We first give a loose characterization of effects that can be accurately estimated and then argue that they are meaningful effects.

%An effect $\xi$ can be accurately estimated if and only if $-\xi$ can be accurately estimated. For brevity, we consider only $\xi$ and extension of results obtained to $-\xi$ is straightforward.

\begin{itemize}

\item[1.] For the $q$ variables in APC arrangement in $\mathbf{X}_1'$ of the standardized model (\ref{m3}), let 
\beq
\xi'(\mathbf{w})=w_1\beta_1'+w_2\beta_2'+\dots+w_q\beta_q' \nonumber %\label{g.eff.s}
\eeq
be a group effect. Its minimum-variance unbiased linear estimator is
\beq
\hspace{0.05in}\hat{\xi}'(\mathbf{w})=w_1\hat{\beta}_1'+w_2\hat{\beta}_2'+\dots+w_q\hat{\beta}_q'.
 \nonumber  %\label{g.eff.s.est}
\eeq
By Theorem \ref{thm1}, the average group effect  ${\xi}_a$ in (\ref{ave.eff}) is accurately estimated as $var(\hat{\xi}_a)$ is substantially smaller than $\sigma^2$.
Since $var(\hat{\xi}'(\mathbf{w}))$ is a continuous function of $\mathbf{w}$, effects ${\xi}'(\mathbf{w})$ in a small neighbourhood of  ${\xi}_a$,
\beq
{\cal N}_a= \{ {\xi}'(\mathbf{w}): ||\mathbf{w}-\mathbf{w}_a||< \delta_1  \} \label{nhood}
\eeq
where $\delta_1$ is a small positive constant and $\mathbf{w}_a=\frac{1}{q}\bold{1}_q$ is the weight vector of $\xi_a$, can also be accurately estimated. Incidentally, there are group effects that can be accurately estimated when variables in $\mathbf{X}_1'$ are not in an APC arrangement, but these effects would be difficult to characterize. The APC arrangement made the simple characterization (\ref{nhood}) possible.

\item[2.] For the $q$ variables in APC arrangement in $\mathbf{X}_1$ of the unstandardised model (\ref{m2}),  the variability weighted average ${\xi}_W$ in (\ref{var.ave}) is accurately estimated by $\hat{\xi}_W$ in (\ref{est2}) as  $var(\hat{\xi}_W)$  is substantially smaller than $\sigma^2$. Other effects  $\xi(\mathbf{w})$ that can be accurately estimated are in a neighbourhood of ${\xi}_W$
\beq
{\cal N}_W= \{ {\xi}(\mathbf{w}): ||\mathbf{w}-\mathbf{w}^*||< \delta_2 \}, \label{nhood2}
\eeq 
where $\delta_2$ is a small positive constant.
An alternative way to characterize ${\cal N}_W$ is to use ${\cal N}_a$ as follows. Let $\xi(\mathbf{w})=\kappa\times \xi'(\mathbf{w}')$ where $\kappa=\sum^q_{i=1}|w_is_i^{-1}|$ and $\xi'(\mathbf{w}')$ is a group effect for $\mathbf{X}_1'$ in the corresponding standardized model with weights $\mathbf{w}'=(w_1', w_2', \dots, w_q')^T$ where $w_i'=w_is_i^{-1}/\kappa$. Usually, $\kappa$ is small as $s_i$ is in general much larger than $w_i$. Thus, $\xi(\mathbf{w})$ can be accurately estimated if $\xi'(\mathbf{w}')$ can be accurately estimated, so 
\beq
{\cal N}_W= \{ {\xi}(\mathbf{w}): \mbox{${\xi}(\mathbf{w})$ such that the corresponding $\xi'(\mathbf{w}')\in {\cal N}_a$}  \}. \nonumber % \label{nhood2}
\eeq

\end{itemize}

{\bf Remark [a]} Set ${\cal N}_a$ in (\ref{nhood}) is also the set of practically important and meaningful group effects for variables in $\mathbf{X}_1'$ in that $\mathbf{w}$ values in the neighbourhood of $\mathbf{w}_a$ represent the most probable changes of the variables in $\mathbf{X}_1'$. Two extreme examples illustrate this point. [1] Effect $\beta_1'\notin {\cal N}_a$ as its weight vector $\mathbf{w}_1=(1,0,\dots,0)$. It represents the group impact on response when $x_1'$ increases by $1$ unit but the other variables do not change. [2] Effect ${\xi}_a$ has $\mathbf{w}_a=(1/q,1/q,\dots,1/q)$, so ${\xi}_a\in {\cal N}_a$. It represents the group impact when all variables increase by $(1/q)$th of a unit. With strong positive correlations and in standardized units, the variables are likely to increase at the same time and in similar amounts. So ${\xi}_a$ is practically important and meaningful whereas $\beta_1'$ is not. In fact, estimating $\beta_1'$ alone amounts to extreme extrapolation and $\beta_1'$ by itself is neither meaningful nor interpretable as one cannot just increase $x_1'$ by 1 unit while holding other variables constant under strong correlations among variables. Another example showing individual parameters are not meaningful is the extreme case of perfect correlation with ${x}_1'=\dots={x}_q'={x}'$. Let $c=\beta_1'+\dots+\beta_q'$. Then, the collective impact of these $q$ variables on the response is $c{x}'$. There are infinitely many sets of  $\beta_i'$ that sum up to $c$. The data $(\mathbf{X}, \mathbf{y})$ contains no information on which set is in the true model. In this sense, it contains no information about the individual $\beta_i'$. Similarly, the data contains little information about the individual $\beta_i'$ when the level of multicollinearity is high. The large variances of the least squares estimators for the $\beta_i'$ are warnings for this lack of information which is always a problem regardless the method of regression used. With this understanding, we should focus on estimating $c$, or equivalently $\xi_a=c/q$, and group effects in ${\cal N}_a$. For the strongly correlated variables in  $\mathbf{X}_1$ in the unstandardised model, a group effect is meaningful if and only if the corresponding effect in the standardized model is meaningful. Thus, ${\cal N}_W$ is the set of meaningful group effects for these variables.

\vspace{0.1in}

{\bf Remark  [b]} Set ${\cal N}_a$ leads to the following geometric characterization of linear combinations $c_1\beta_1'+c_2\beta_2'+ \dots+c_q\beta_q'$ that can be accurately estimated for the standardized model (\ref{m3}). A linear combination can be expressed as $c_t\xi'(\mathbf{w})$ where $c_t=\sum^q_{i=1} |c_i|$ and $\mathbf{w}={c_t}^{-1}(c_1, c_2,\dots,c_q)^T$. Its minimum-variance unbiased linear estimator is $c_t\hat{\xi}'(\mathbf{w})$, so it can be accurately estimated when $var(c_t\hat{\xi}'(\mathbf{w}))=c_t^2 var(\hat{\xi}'(\mathbf{w}))$ is small relative to $\sigma^2$. This happens under one of the following two conditions: ($i$) $\xi'(\mathbf{w}) \in {\cal N}_a$ and $c_t$ is not too large, or ($ii$) $\xi'(\mathbf{w}) \notin {\cal N}_a$ but $c_t$ is very small. These two conditions and ${\cal N}_a$ imply that in the 2-dimensional case where $q=2$, points $(c_1, c_2)$ representing linear combinations that can be accurately estimated form a band centred around the line $c_1=c_2$. In higher dimensions where $q>2$, they form a hyper-cylinder centred around the line $c_1=c_2=\dots=c_q$. This observation will be used for discussing prediction accuracy in Section 3.3.

%3333333333333333333333333333333333333333333333333333333333333333333333333333333333333333333333333333333333333333333333333333

\section{Group approach to the least squares regression}

In this section, we present the group approach through examples. In particular, we present a group effect based analysis on the prediction accuracy of the least squares estimated model. For simplicity, we use a small model (\ref{m10}) throughout this section but there is no loss of generality as similar results can be obtained for models of any size. Consider model (\ref{m10}) with 6 predictor variables in 4 groups $\mathbf{X}_1=[\mathbf{x}_1,\mathbf{x}_2]$, $\mathbf{X}_2= [\mathbf{x}_3,\mathbf{x}_4]$, $\mathbf{X}_3=[\mathbf{x}_5]$ and $\mathbf{X}_4=[\mathbf{x}_6]$,
\begin{equation}
\mathbf{y}=\beta_0\mathbf{1}_n+\mathbf{X}_1\boldsymbol{\beta}_1+\mathbf{X}_2\boldsymbol{\beta}_2 +\mathbf{X}_3\boldsymbol{\beta}_3 +\mathbf{X}_4\boldsymbol{\beta}_4 +\boldsymbol{\varepsilon}, \label{m10}
\end{equation}
where $\beta_0=3$, $\boldsymbol{\beta}_1=(\beta_1, \beta_2)^T=(0,0)^T$, $\boldsymbol{\beta}_2=(\beta_3, \beta_4)^T=(1,2)^T$, $\boldsymbol{\beta}_3=\beta_5=0$, $\boldsymbol{\beta}_4=\beta_6=3$ and $\boldsymbol{\varepsilon}$ is the $n$-variate normal random error with $\sigma^2=1$. We use 6 i.i.d. $n$-variate standard normal random vectors $\mathbf{z}_i$ and three parameters $(w_1,w_2,\gamma)$ to generate the 6 variables as follows so that $\mathbf{X}_1$ and $\mathbf{X}_2$ are, respectively, strongly correlated groups,
\bey
& & \mathbf{x}_1=\mathbf{z}_1, \hspace{0.1in}  \mathbf{x}_2=\gamma[w_1\mathbf{z}_1+(1-w_1)\mathbf{z}_2]; \nonumber \\
& & \mathbf{x}_3=\mathbf{z}_3, \hspace{0.1in}  \mathbf{x}_4=\gamma[w_2\mathbf{z}_3+(1-w_2)\mathbf{z}_4];\label{expl} \\
& & \mathbf{x}_5=\mathbf{z}_5, \hspace{0.1in} \mathbf{x}_6=\gamma \mathbf{z}_6. \nonumber
\eey
%The theoretical non-zero correlation coefficients are:
%\bey
%& & \sigma_{12}=\sigma_{21}=w_1[w_1^2+(1-w_1)^2]^{-1/2} ; \nonumber \\
%& &  \sigma_{34}=\sigma_{43}=w_2[w_1^2+(1-w_2)^2]^{-1/2}. \nonumber
%\eey
%For our simulation study, we use observed values of the six $\mathbf{x}_i$ variables, so the sample correlation coefficients differ somewhat from the theoretical value.
We set $n=12$, $w_1=0.7$, $w_2=0.8$ and $\gamma=2$. Matrix $\mathbf{X}_d=[\mathbf{x}_1,\mathbf{x}_2,\dots,\mathbf{x}_6]$ containing numerical values of the 6 variables randomly generated using (\ref{expl}) is %saved in xsave1 matrix in R
given in ``R display 1'' in the Supplementary Material. 
The full design matrix is $\mathbf{X}=[\mathbf{1}_n, \mathbf{X}_d]$. 
Table \ref{table0} contains the correlation matrix of the 6 variables in $\mathbf{X}_d$ which shows strong within-group correlations but weak between-group correlations. We consider only the unstandardised model in this section. An example of the standardized model is given in Section 4.

\renewcommand{\baselinestretch}{1.0}
\begin{table}
\caption{\label{table0}Correlation coefficients of the 6 variables in $\mathbf{X}_d=[\mathbf{x}_1,\mathbf{x}_2,\dots,\mathbf{x}_6]$ of model (\ref{m10})}
\centering
%\fbox{ %
\begin{tabular}{rrrrrrr}
& $\mathbf{x}_1$ & $\mathbf{x}_2$ & $\mathbf{x}_3$ & $\mathbf{x}_4$ & $\mathbf{x}_5$ & $\mathbf{x}_6$ \\
$\mathbf{x}_1$ & 1.00 & 0.90 &-0.34 & -0.34 & -0.06 & 0.14\\
$\mathbf{x}_2$ & 0.90 & 1.00 &-0.27 & -0.20 & -0.25 & 0.38\\
$\mathbf{x}_3$ & -0.34&-0.27 & 1.00 & 0.96  & -0.41 & -0.53\\
$\mathbf{x}_4$ & -0.34&-0.20 & 0.96 & 1.00  & -0.49 & -0.44\\
$\mathbf{x}_5$ & -0.06&-0.25 &-0.41 & -0.49 & 1.00  &  0.03\\
$\mathbf{x}_6$ & 0.14 & 0.38 &-0.53 & -0.44 & 0.03  &  1.00\\
\end{tabular}%}
\end{table}
\renewcommand{\baselinestretch}{1.6}

\subsection{Group approach to estimation and inference}

For a group of strongly correlated variables in an unstandardised model, the group approach studies only meaningful group effects in the neighbourhood of its variability weighted average (\ref{nhood2}). To compare such effects with effects not in the neighbourhood, we consider the following six effects for model (\ref{m10}):

\vspace{0.1in}

\noindent \hspace*{0.3in} 1. $\xi_1=w^*_{11}\beta_1+w^*_{12}\beta_2$: variability weighted average for group $\mathbf{X}_1$.\\
\noindent \hspace*{0.3in} 2. $\xi_2=w^*_{21}\beta_3+w^*_{22}\beta_4$: variability weighted average for group $\mathbf{X}_2$.\\
\noindent \hspace*{0.3in} 3. $\xi_3=\frac{1}{2}(\beta_1-\beta_2)$: half difference effect for  group $\mathbf{X}_1$.\\
\noindent \hspace*{0.3in} 4. $\xi_4=\frac{1}{2}(\beta_5-\beta_6)$: half difference effect between $\mathbf{x}_5$ and $\mathbf{x}_6$.\\
\noindent \hspace*{0.3in} 5. $\xi_5=\frac{1}{2}(\beta_3+\beta_4)$: average group effect for group $\mathbf{X}_2$.\\
\noindent \hspace*{0.3in} 6. $\xi_6=(w^*_{21}-\delta)\beta_3+(w^*_{22}+\delta)\beta_4$: an effect in the neighbourhood of $\xi_2$.

\vspace{0.1in}
\noindent The weight vector for $\xi_1$ is  $(w^*_{11}, w^*_{12})=(0.42847, 0.57152)$ and that for $\xi_2$ is $(w_{21}^*, w_{22}^*)=(0.39177, 0.60822)$. The exact values of the six effects $\xi_i$ are $0, 1.60822, 0, -1.5, 1.5, 1.65822$, respectively.
Table \ref{table1} gives the means and variances of 1000 minimum-variance unbiased linear estimates for these six group effects and the six parameters of model (\ref{m10}). We used the same design matrix $\mathbf{X}_d$ in ``R display 1''  and model (\ref{m10}) to randomly generate 1000 $\mathbf{y}$'s. Each estimate was computed by using one of the 1000 $(\mathbf{X}_d, \mathbf{y})$ pairs.

\begin{table}
\caption{\label{table1} Mean and variance of 6 estimated group effects and 6 estimated individual effects based on 1000 simulated values. }
\centering
\fbox{%
%\begin{tabular}{*{6}{c}}
\begin{tabular}{ccc|ccc}
Effect &       Mean      & Variance  &  Effect  &    Mean     & Variance \\ \hline
$\xi_1$   &  0.01009    &0.02643 &$\beta_1$ & 0.01604  &2.16007 \\
$\xi_2$   &  1.61319    &0.03534 &$\beta_2$ & 0.00526  &1.37544 \\
$\xi_3$   &  0.05936    &1.68234 &$\beta_3$ & 1.01535  &1.66295 \\
$\xi_4$   & -1.49600    &0.08343 &$\beta_4$ & 1.98636  &0.82435 \\
$\xi_5$   &  1.50585    &0.06974 &$\beta_5$ & 0.00688  &0.13240\\
$\xi_6$   &  1.66424    &0.05442 &$\beta_6$ & 3.00181  &0.14773
\end{tabular}}
\end{table}

Table \ref{table1} shows $\xi_1$ and $\xi_2$ are accurately estimated with very small variances relative to the error variance $\sigma^2=1$. Effect $\xi_3$ is the half difference effect for $\mathbf{X}_1$ which is not in the neighbourhood of $\xi_1$ as its weight vector $(0.5,-0.5)$ is not close to $(w^*_{11}, w^*_{12})$, so it is poorly estimated with a large variance. But since  $\xi_3$ measures the expected change in the response when $x_1$ increases by half a unit and $x_2$ decreases by half a unit at the same time which is unlikely to occur given the strong positive correlation between $x_1$ and $x_2$, it is not a practically meaningful effect, so we are not interested in $\xi_3$ and thus not concerned that it cannot be accurately estimated. Effect $\xi_4$ is also a half difference effect but for weakly correlated $\mathbf{x}_5$ and $\mathbf{x}_6$. It is accurately estimated. Effect $\xi_5$ is the average group effect of $\mathbf{X}_2$. It is accurately estimated as it is in the neighbourhood of the variability weighted average effect $\xi_2$. Effect $\xi_6$ of $\mathbf{X}_2$ will be in the neighbourhood of $\xi_2$ when $\delta$ is small. For the $\xi_6$ in Table \ref{table1}, $\delta=0.05$, so it is accurately estimated. Parameters $\beta_1, \beta_2,\beta_3$ and $\beta_4$ for the two strongly correlated groups are poorly estimated but $\beta_5$ and $\beta_6$ are accurately estimated. In real applications, there is only one response vector $\mathbf{y}$ and thus only one estimated value $\hat{\xi}(\mathbf{w})=w_1\hat{\beta}_1+w_2\hat{\beta}_2+\dots+w_6\hat{\beta}_6$ for $\xi(\mathbf{w})$. To assess whether $\hat{\xi}(\mathbf{w})$ is accurate, we may use the estimated variance  $\widehat{var}(\hat{\xi})$ which can be computed by using (\ref{evar}) with $\mbox{\boldm $x$}_+= (0,w_1,\dots,w_6)$. 

To test hypotheses or construct confidence intervals for $\xi(\mathbf{w})$, we use
\beq
T=\frac{\hat{\xi}(\mathbf{w})-\xi(\mathbf{w})}{\sqrt{\widehat{var}(\hat{\xi})}} \label{ttest}
\eeq
which has a $t_{n-7}$ distribution under the null hypothesis. To summarize, for strongly correlated variables in an unstandardised model, meaningful group effects in the neighbourhood of the variability weighted average are accurately estimated. For variables not strongly correlated with others, estimates for their parameters and effects are not affected by multicollinearity and are accurate. Inference for group effects can be done by using the $t$ statistic in (\ref{ttest}).

\subsection{Group approach to variable selection}

Traditional methods of variable selection such as all subsets regression and stepwise selection allow variables to be selected one at a time. Multicollinearity creates problems for these methods as often only one variable from a strongly correlated group is selected and different methods may choose very different models.
The group approach does variable selection at the group level so that variables in a group are either all in or all out. We now illustrate this through all subsets regression for model (\ref{m10}). Recall that $\beta_0=2$, $\boldsymbol{\beta}_1=(0,0)^T$, $\boldsymbol{\beta}_2=(1,2)^T$, $\boldsymbol{\beta}_3=0$ and $\boldsymbol{\beta}_4=3$, so the ``true model'' is the 3-variable model:
\begin{equation}
\mathbf{y}=\beta_0\mathbf{1}_n+{\beta}_3\mathbf{x}_3+{\beta}_4\mathbf{x}_4 +{\beta}_6\mathbf{x}_6 +\boldsymbol{\varepsilon}. \nonumber \label{m12}
\end{equation}
There are $2^6-1=63$ non-empty models with at least one variable. Among these, 15 are what we call ``group models'' where $x_1$ and $x_2$ are in or out at the same time, and $x_3$ and $x_4$ are in or out at the same time. Using R package ``leaps'' by Lumley and Miller (2017), we performed all subsets regression with the adjusted $R^2$ criterion 100 times using 100 sets of simulated data from model (\ref{m10}). In each run, the model with the highest adjusted $R^2$ value among all 63 models is the choice of the traditional all subsets regression and that among the 15 group models is the choice of the group approach to all subsets regression.  Table \ref{table01} gives a partial summary of the results of the 100 runs. A full table containing all 21 models chosen at least once and ``R display 2'' containing a sample run may be found in the Supplementary Material.  We make the following observations based on results in Table \ref{table01}:

\renewcommand{\baselinestretch}{1.0}
\begin{table}
\caption{\label{table01} Percentage of times a model is chosen by the traditional all subsets regression (Pct$_1$) and group approach to all subsets regression (Pct$_2$). There are 21 models that were chosen at least once by either method and 12 of these are listed in this table.}
\centering
%\fbox{ %
\begin{tabular}{|cccc|} \hline
Model                            & Group model? & Pct$_1$  & Pct$_2$  \\ \hline
$x_3, x_4, x_6$                  & Yes            & 14\%                    & 45\% \\
$x_3, x_4, x_5, x_6$             & Yes            & 3\%                     & 22\% \\
$x_1, x_2, x_3, x_4, x_6$        & Yes            & 2\%                     & 18\% \\
$x_1, x_2, x_3, x_4, x_5, x_6$   & Yes            & 5\%                     & 15\% \\
$x_4, x_6$                       & No             & 18\%                    & 0\% \\
$x_4, x_5, x_6$                  & No             & 11\%                    & 0\% \\
$x_1, x_5, x_6$                  & No             & 4\%                     & 0\% \\
$x_1, x_3, x_4, x_6$             & No             & 2\%                     & 0\% \\
$x_1, x_2, x_4, x_6$             & No             & 8\%                     & 0\% \\
$x_2, x_4, x_6$                  & No             & 2\%                     & 0\% \\
$x_2, x_3, x_4, x_6$             & No             & 6\%                     & 0\% \\ 
$x_1, x_2, x_4, x_5, x_6$        & No             & 4\%                     & 0\% \\  \hline
%$x_1, x_3, x_4, x_5, x_6$        & No             & 5\%                     & 0\% \\
%$x_1, x_4, x_5, x_6$             & No             & 1\%                     & 0\% \\
%$x_2, x_3, x_4, x_5, x_6$        & No             & 1\%                     & 0\% \\
%$x_2, x_4, x_5, x_6$             & No             & 5\%                     & 0\% \\
%$x_1, x_2, x_3, x_6$             & No             & 1\%                     & 0\% \\
%$x_1, x_2, x_3, x_5, x_6$        & No             & 3\%                     & 0\% \\
%$x_3, x_6$                       & No             & 1\%                     & 0\% \\
%$x_3, x_5, x_6$                  & No             & 3\%                     & 0\% \\
%$x_2, x_3, x_6$                  & No             & 1\%                     & 0\% \\ \hline
\end{tabular}%} 
\end{table}

\begin{itemize}

\item[1.] In the 100 simulation runs, 4 of the 15 group models (roughly 1/4) were chosen at least once by the group approach, but 21 of 63 models (or 1/3) were chosen by the traditional method, so the group approach is more stable in its selection.
The true model containing $\{x_3, x_4, x_6\}$ was chosen 45\% of the time by the group approach but only 14\% of the time by the traditional method, so the group approach is also more accurate.
\item[2.] When the traditional and group approach picked different models, the adjusted $R^2$ values of their chosen models typically differ by less than 1\% (see the run in ``R display 2" for an example). This shows the group approach is competitive  in terms of the adjusted $R^2$ value of the chosen model.
\item[3.] All 4 models that have been picked by the group approach at least once contain all relevant variables (variables with non-zero parameters). In contrast, 80\% of the models picked by the traditional method have missed at least one relevant variable; for example, for the run in ``R display 2'', $x_4$ is in but $x_3$ is out even though the parameter $\beta_3\neq 0$.

\end{itemize}

We also performed variable selection for a different version of model (\ref{m10}) where $\boldsymbol{\beta}_2=(1,0)$. In this case, the true model contains only $\{x_3, x_6\}$ which cannot be recovered by the group approach as it is not a group model. The group approach picked model $\{x_3, x_4, x_6\}$ with the highest frequency.
The above example involves all subsets regression. We may apply the group approach with a different model selection method such as backward selection.
Numerical results show that under the group approach, different methods are more consistent in that they are more likely to select the same model.

%It should be noted that the idea of variable selection by groups was first considered by Yuan and Lin (2006). There are, however, two differences in our use of this idea. Firstly, Yuan and Lin (2006) is concerned about groups of variables derived from the same factors, whereas here we are concerned about groups of strongly correlated variables. Secondly, Yuan and Lin (2006) considers variable selection in the high dimensional setting with penalized least squares methods, whereas here we consider traditional variable selection methods for the least squares regression in the classical ``$n>p$'' setting. 

\subsection{Group approach to prediction accuracy analysis}

Multicollinearity often leads to poor predictions but it is known that accurate predictions may be achieved in an area of the variable space. This area is usually expressed through an approximate linear constraint involving all variables; see, for example, (9.1) on page 286 and remarks about prediction accuracy on page 290 in Montgomery, Peck and Vining (2012). However, such a constraint provides only a vague description of the area where accurate predictions can be achieved. We now take the group approach to characterize such an area and also address the misconception mentioned in the introduction. 

Consider the expected response at {\boldm $x$} $=(x_1,\dots,x_6)$ under model (\ref{m10}),
\begin{equation}
E(y|\mbox{{\boldm $x$}}) = \beta_0+x_1\beta_1+x_2\beta_2+x_3\beta_3+x_4\beta_4+x_5\beta_5+x_6\beta_6, \label{expected}
\end{equation}
where $\beta_j$ are the unknown parameters and {\boldm $x$} is a row vector containing values of the 6 predictor variables. The predicted value for $E(y|\mbox{{\boldm $x$}})$ by the least squares estimated model is
\begin{equation}
\hat{y}= \hat{\beta}_0+x_1\hat{\beta}_1+x_2\hat{\beta}_2+x_3\hat{\beta}_3+x_4\hat{\beta}_4+x_5\hat{\beta}_5+x_6\hat{\beta}_6, \label{estimated}
\end{equation}
where $\hat{\beta}_j$ are the least squares estimates of $\beta_j$. 
%To find the area of  {\boldm $x$} in which $\hat{y}$ is an accurate predictor for $E(y|\mbox{{\boldm $x$}})$, we use the standardized version of model % (\ref{m10}).
Let $\mathbf{y}'=\mathbf{y}-\bar{y}$ be the centred version of $\mathbf{y}$ and $\mathbf{x}_i'$ be the standardized version of $\mathbf{x}_i$ in model (\ref{m10}). Then,
\begin{equation}
\mathbf{y}'=\mathbf{x}_1'{\beta}_1'+\mathbf{x}_2'{\beta}_2' +\mathbf{x}_3'{\beta}_3' +
\mathbf{x}_4'{\beta}_4'+\mathbf{x}_5'{\beta}_5' +\mathbf{x}_6'{\beta}_6'+\boldsymbol{\varepsilon} \label{m30}
\end{equation}
is the standardized version of model (\ref{m10}). Let $\hat{\beta}_i'$ be the least squares estimates for parameters of (\ref{m30}). They are related to  $\hat{\beta}_j$ for (\ref{m10}) as follows,
\beq
\hat{\beta}_0=\bar{y}-\sum^6_{i=1}\bar{x}_i\hat{\beta}_i'/s_i \mbox{\hspace{0.1in} and \hspace{0.1in}} \hat{\beta}_i=\hat{\beta}_i'/s_i
\mbox{\hspace{0.1in} for $i=1,2,\dots,6$}, \label{coeffs}
\eeq
where $\bar{x}_i$ and $s_i$ are defined just above equation (\ref{std.x}).  By (\ref{estimated}) and (\ref{coeffs}), 
%$\hat{y}$ can be expressed in terms $\hat{\beta}_i'$ as follows,
\beq
\hat{y}=(\bar{y}-\sum^6_{i=0}\bar{x}_i\hat{\beta}_i'/s_i) + x_1(\hat{\beta}_1'/s_1) + \dots + x_6(\hat{\beta}_6'/s_6). \label{temp1}
\eeq
Define the ``standardized'' version of $\mbox{\boldm $x$}$, $\mbox{\boldm $x$}'=(x_1',x_2', \dots,x_6')$, as
\beq
x_i'=\frac{x_i-\bar{x}_i}{s_i} \hspace{0.2in} \mbox{for $i=1,2,\dots,6$}. \label{trans}
\eeq
Using (\ref{temp1}) and (\ref{trans}), we obtain an expression of $\hat{y}$ in terms of $\hat{\beta}_i'$ and $x_i'$,
\beq
\hat{y}= \bar{y}+(x_1'\hat{\beta}_1'+x_2'\hat{\beta}_2')+(x_3'\hat{\beta}_3'+x_4'\hat{\beta}_4')+(x_5'\hat{\beta}_5')+(x_6'\hat{\beta}_6'). \label{temp2}
\eeq
Since $\hat{y}$ is unbiased for $E(y|\mbox{{\boldm $x$}})$, taking expectation on both sides of (\ref{temp2}) shows that $E(y|\mbox{{\boldm $x$}})$ is the sum of the expectations of the 5 terms in the right-hand side of (\ref{temp2}). Thus, if all 5 terms accurately estimate their respective expectations, then $\hat{y}$ is an accurate estimate of $E(y|\mbox{{\boldm $x$}})$. The $\bar{y}$ accurately estimates $E(y)$. Also, $\hat{\beta}_5'$ and $\hat{\beta}_6'$ are accurate estimators as they are for parameters of variables not strongly correlated with others, so $x_5'\hat{\beta}_5'$ and $x_6'\hat{\beta}_6'$ accurately estimate their expected values. Since $x_1'$ and $x_2'$ are strongly correlated, by Remark [b] of Section 2.3, $(x_1'\hat{\beta}_1'+x_2'\hat{\beta}_2')$ accurately estimates its expectation $(x_1'\beta_1'+x_2'\beta_2')$  if $(x_1', x_2')\in {\mathcal C}_1'$ where ${\mathcal C}_1'$ is a band centred around the line $x_1'=x_2'$. Similarly, $(x_3'\hat{\beta}_3'+x_4'\hat{\beta}_4')$ accurately estimates $(x_3'\beta_3'+x_4'\beta_4')$  if $(x_3', x_4')\in {\mathcal C}_2'$ where ${\mathcal C}_2'$ is a band centred around the line $x_3'=x_4'$. Thus, the region of $\mbox{\boldm $x$}'$ over which $\hat{y}$ is an accurate estimation for $E(y|\mbox{{\boldm $x$}})$ is 
\beq
{\mathcal R}_{FP}'={\mathcal C}_1'\times {\mathcal C}_2'\times \mathbb{R}^2   \label{fpr1}
\eeq
where the $\mathbb{R}^2$ represents no restrictions on variables $x_5'$ and $x_6'$ as they are not strongly correlated with other variables. We call the region in (\ref{fpr1}) the feasible prediction region for the least squares estimated model (\ref{estimated}). 
%The region in (\ref{fpr1}) is expressed in the standardized variable $\mbox{\boldm $x$}'$. 
In terms of the unstandardised variable $\mbox{\boldm $x$}$, the feasible prediction region is
\beq
{\mathcal R}_{FP}=\{\mbox{\boldm $x$} : \mbox{\boldm $x$} \hspace{0.05in} \mbox{such that its corresponding} \hspace{0.05in} \mbox{\boldm $x$}' \in {\mathcal R}_{FP}'\}. \nonumber \label{fpr2}
\eeq
In simple terms, the feasible prediction region is the region in the predictor variable space where each group of strongly correlated variables in their APC arrangement are approximately equal after standardization (\ref{trans}). The least squares estimated model gives accurate predictions over this region. 

The variance of a predicted value ${var}(\hat{y})$ is estimated by
\beq
 \widehat{var}(\hat{y})=\hat{\sigma}^2\mbox{\boldm $x$}_+(\mathbf{X}^T\mathbf{X})^{-1}\mbox{\boldm $x$}_+^T,\label{evar}
\eeq
where $\mbox{\boldm $x$}_+=(1, \mbox{\boldm $x$})= (1,x_1,\dots,x_6)$ and $\hat{\sigma}^2$ is the mean squared error. The accuracy of $\widehat{var}(\hat{y})$ depends only on the accuracy of $\hat{\sigma}^2$ as an estimator for $\sigma^2$ which is known to be good and unaffected by multicollinearity. Thus, $\widehat{var}(\hat{y})$ is in general accurate and unaffected by the multicollinearity in the data.

To illustrate ${\mathcal R}_{FP}$, we make predictions using the least squares estimated model (\ref{estimated}) and the ridge regression at the following three points: \vspace{0.1in}

\noindent \hspace*{0.8in} $\mbox{\boldm $x$}_1=(0.60413, 0.75045, 0.00328, 0.21336, 1, 2),$ \\
\hspace*{0.8in}  $\mbox{\boldm $x$}_2=(0.93025, 1.27245, 0.75025, 1.48901, 1, 2),$\\
\hspace*{0.8in}  $\mbox{\boldm $x$}_3=( 1.58247,  1.18545, 0.75025, 3.11257, 1, 2).$     \vspace{0.1in}

\noindent Using (\ref{trans}) and $\mathbf{X}_d$ in ``R display 1'' in the Supplementary Material, we find the standardized versions of these $\mbox{\boldm $x$}_i$, and they are

 \vspace{0.1in}

\noindent \hspace*{0.8in} $\mbox{\boldm $x$}_1'=(0,0,0,0,*,*),$ \\
\hspace*{0.8in}  $\mbox{\boldm $x$}_2'=(0.10, 0.12, 0.20, 0.22, *, *),$\\
\hspace*{0.8in}  $\mbox{\boldm $x$}_3'=( 0.30, 0.10,  0.20, 0.50, *,*).$

 \vspace{0.1in}

\noindent where the standardized values of $x_5$ and $x_6$ are not shown as they are irrelevant. From the standardized values of the first four variables which are in strongly correlated groups, we see that $\mbox{\boldm $x$}_1$ is at the centre of ${\mathcal R}_{FP}$ as $\mbox{\boldm $x$}_1'$ is at the centre of ${\mathcal R}_{FP}'$; $\mbox{\boldm $x$}_2$ is also in ${\mathcal R}_{FP}$ as $\mbox{\boldm $x$}_2'$ is in  ${\mathcal R}_{FP}'$ ($0.1\approx 0.12$ and $0.2\approx 0.22$), but $\mbox{\boldm $x$}_3$ is not in ${\mathcal R}_{FP}$ as $\mbox{\boldm $x$}_3'$ is not in  ${\mathcal R}_{FP}'$ ($0.30 \not\approx 0.10$ and $0.20 \not\approx 0.50$).

\begin{table}
\caption{\label{table2} Comparison of the least squares and Ridge regression predictors for $E(y)$ at predictor vector values $\mbox{\boldm $x$}_1$, $\mbox{\boldm $x$}_2$ and $\mbox{\boldm $x$}_3$ in terms of estimated bias (in absolute value) and MSE based on 1000 simulated values of each predictor.} %Model setting: $(w_1,w_2)=(0.999,0.999)$.}
\centering
\begin{tabular}{|c|c|c c|c c|} \hline
$\mbox{\boldm $x$}$ & Exact & \multicolumn{2}{c|}{Least squares} &\multicolumn{2}{c|}{Ridge regression} \\ \cline{3-6}
values           & $E(y)$  &  Bias       & MSE        & Bias        & MSE  \\ \hline
$\mbox{\boldm $x$}_1$ & 9.43000  & 0.02184 & 0.78324  &  0.29714 & 0.83051\\
$\mbox{\boldm $x$}_2$ & 12.72829 & 0.03562 & 1.41920  &  0.42563 & 1.55798 \\
$\mbox{\boldm $x$}_3$ & 15.97541 & 0.10922 & 9.91271  &  1.02438 & 7.84208 \\ \hline
\end{tabular}
\end{table}

Table \ref{table2} contains the bias and MSE of the least squares predictor (\ref{estimated}) and the ridge regression predictor based on 1000 simulated values of the two predictors computed by using the same design matrix $\mathbf{X}_d$ but 1000 different $\mathbf{y}$ values simulated using model (\ref{m10}).
The least squares predictor has small bias at all three $\mbox{\boldm $x$}_i$ points as it is unbiased. Its MSE is small at $\mbox{\boldm $x$}_1$ and $\mbox{\boldm $x$}_2$ but large at $\mbox{\boldm $x$}_3$ because $\mbox{\boldm $x$}_1$ and $\mbox{\boldm $x$}_2$ are in ${\mathcal R}_{FP}$ but $\mbox{\boldm $x$}_3$ is not.  The ridge regression predictions were computed by using R package ``glmnet'' by Friedman \emph{et al.} (2017)
with the optimal $\lambda$ value in $(0.01, 1000)$. It has bigger bias than the least squares predictor  at all three points. At $\mbox{\boldm $x$}_1$ and $\mbox{\boldm $x$}_2$, its MSE is larger than that of the least squares predictor. At $\mbox{\boldm $x$}_3$, its MSE is smaller but is large in absolute terms. We have compared the two predictors using other examples and observed the same behaviour: at an $\mbox{\boldm $x$} \in {\mathcal R}_{FP}$, both predictors are accurate but the least squares predictor is more accurate with smaller bias and smaller MSE. Outside ${\mathcal R}_{FP}$, the ridge regression predictor has a smaller MSE but a larger bias, and neither estimator is very accurate. 

The misconception that the ridge regression gives more accurate predictions than the least squares regression was based on comparing prediction accuracy  outside ${\mathcal R}_{FP}$ which was unknowingly done as the concept of feasible prediction region ${\mathcal R}_{FP}$ was previously unavailable. From (\ref{temp2}), we see that making a prediction amounts to estimating a set of group effects. Making predictions over ${\mathcal R}_{FP}$ involves estimating meaningful effects, but doing so outside ${\mathcal R}_{FP}$ involves estimating effects that are not meaningful  (see Remark [a] in Section 2.3). Thus, predictions outside ${\mathcal R}_{FP}$ are also not meaningful, and they should not be used for comparison. When we compare meaningful predictions over ${\mathcal R}_{FP}$, the least squares predictor is more accurate.

Finally, as an example of estimating the variance of the least squares predictor with formula (\ref{evar}), for the 3 points in Table \ref{table2}, the average of 1000 estimates by (\ref{evar}) are 0.72335,  1.38200 and 9.21323, respectively, which match the MSE's in Table \ref{table2} closely.  On the other hand, there is no simple formula for estimating the variance of the ridge regression predictor when $\lambda$ is optimized.

%44444444444444444444444444444444444444444444444444444444444444444444444444444444444444444444444444444444444444444444444444444444444444444
%%%%%%%%%%%%%44444444444444444444444444444444444444444444444444444444444444

\section{Application to Hald cement data}

The Hald cement data has been widely used in the literature to illustrate multicollinearity; see, for example, Draper and Smith (1998). Here, we use this data set to illustrate several points in this paper. The data set contains 13 observations with 4 predictor variables and a response $y$:\\[0.1in]
\hspace*{0.2in} $y$ = heat evolved in calories per gram of cement; \\
\hspace*{0.2in} $x_1$ = amount of tricalcium aluminate;\\
\hspace*{0.2in} $x_2$ = amount of tricalcium silicate;\\
\hspace*{0.2in} $x_3$ = amount of tetracalcium alumino ferrite;\\
\hspace*{0.2in} $x_4$ = amount of dicalcium silicate. \\[0.10in]
We first illustrate the APC arrangement of a group of strongly correlated variables.
In Table \ref{h1}, the correlation matrix on the left is that of the 4 predictor variables in the Hald cement data. It shows that there are two strongly correlated groups $\{x_1,x_3\}$ and $\{x_2,x_4\}$ with negative correlation within each group, so $\{x_1,-x_3\}$ and $\{x_2,-x_4\}$ are their APC arrangements. For convenience, we rename the variables so that $x_1$ is still the same but the old $-x_3$ is now called $x_2$, the old $x_2$ now called $x_3$, and the old $-x_4$ now called $x_4$. The correlation matrix of the renamed variables is on the right of Table \ref{h1}. The strongly correlated groups are now $\{x_1,x_2\}$ and $\{x_3,x_4\}$, both in APC arrangement, and there are no strong correlations between variables from different groups.

\begin{table}
\caption{\label{h1} Correlations of original Hald cement data (left) and renamed data (right)}
\centering
%\fbox{ %
\begin{tabular}{rrrrrrrrrrr}
& $\mathbf{x}_1$ & $\mathbf{x}_2$ & $\mathbf{x}_3$ & $\mathbf{x}_4$ &  & & $\mathbf{x}_1$ & $\mathbf{x}_2$ & $\mathbf{x}_3$ & $\mathbf{x}_4$  \\
$\mathbf{x}_1$ & 1.00 & 0.22 &-0.82 & -0.24 &  &$\mathbf{x}_1$ & 1.00 & 0.82 &-0.22 & 0.24\\
$\mathbf{x}_2$ & 0.22 & 1.00 &-0.13 & -0.97 &  &$\mathbf{x}_2$ & 0.82 & 1.00 &0.13 & 0.02\\
$\mathbf{x}_3$ & -0.82 &-0.13 &1.00 & 0.02  &  &$\mathbf{x}_3$ & 0.22 & 0.13 & 1.00 & 0.97\\
$\mathbf{x}_4$ & -0.24 & -0.97 &0.02 & 1.00 &  &$\mathbf{x}_4$ & 0.24 & 0.02 & 0.97 & 1.00\\
\end{tabular} %}
\end{table}

Turning now to the standardized model (\ref{m3}) based on the renamed variables where the matrix $\mathbf{X}'^T\mathbf{X}'$ in (\ref{corr.total}) is just the correlation matrix on the right of Table \ref{h1}. Matrix $\mathbf{R}_{11}$ in (\ref{corr.total}) is the upper-left quarter of this correlation matrix and $\mathbf{R}_{22}$ is the lower-right quarter. A condition used in Theorem \ref{thm1} is that weak correlation between $\mathbf{X}'_1$ and $\mathbf{X}'_2$ leads to small elements in $\mathbf{R}_{12}\mathbf{R}^{-1}_{22}\mathbf{R}_{21}$. To illustrate this, for the present example we have 
\[ \mbox{$\mathbf{R}_{12}\mathbf{R}^{-1}_{22}\mathbf{R}_{21}$}=
\begin{pmatrix}
0.06 & -0.01\\
-0.01 & 0.22
\end{pmatrix}
\]
where the elements are indeed small relative to that of $\mathbf{R}_{11}$ and $\mathbf{R}_{22}$.

\renewcommand{\baselinestretch}{1.0}
\begin{table}
\caption{\label{h2} Estimated parameter values and average group effects for the standardized model (\ref{m3}); $\xi_{a}^1$ is the estimated average group effect for group $\{x_1',x_2'\}$, $\xi_{a}^2$ is that for $\{x_3',x_4'\}$.  }
\centering
\fbox{ %
\begin{tabular}{rrrrrr}
& Estimate & Std. Error & $t$ value &  Pr($>|t|$) \\
${\beta}_1'$ &31.607 & 14.308  &2.209& 0.055 \\
${\beta}_2'$ & -2.261 & 15.788 & -0.143 & 0.889\\
${\beta}_3'$ &27.500 &  36.784 &  0.748 & 0.473 \\
${\beta}_4'$ &8.353   &  38.762 &  0.215 & 0.834\\
$\xi_{a}^1$ &14.673  &  1.456  &  10.072  & 0.000 \\
$\xi_{a}^2$ & 17.927  &   1.571&   11.409 &   0.000
\end{tabular}}
\end{table}
\renewcommand{\baselinestretch}{1.6}
 Table \ref{h2} shows the estimated values of the 4 parameters $\beta'_i$ and the 2 average group effects $\xi_{a}^i$ in (\ref{ave.eff}).
The $\beta'_i$ are poorly estimated with large standard errors due to multicollinearity generated by the two groups of strongly correlated variables. The $t$-test shows they are not significantly different from zero at the 5\% level. The average group effects, on the other hand, are very accurately estimated with small standard errors and are highly significant. The estimated error variance is $\hat{\sigma}^2=2.306^2$, so the (estimated) lower bound for the standard errors of the two group effects from Theorem \ref{thm1} is $\hat{\sigma}/2=1.153$. We see from Table \ref{h2} that the standard errors of the two estimated group effects are quite close to this lower bound. We write the least squares estimated model as
\beq
y'=(31.607x_1'-2.261x_2')_G+(27.500x_3'+8.353x_4')_G,  \label{ee}
\eeq
where the $(\dots)_G$ notation indicates that variables inside each $(\dots)_G$ are strongly correlated. Individual estimated parameter values such as $31.607$ and $-2.261$ inside such brackets should not be used as point estimates as the underlying parameters are not meaningful and thus not estimated; they should only be used to estimate or make inference on meaningful group effects, such as $\xi_{a}^1$ and $\xi_{a}^2$, or make predictions over the feasible prediction region. 

Finally, we demonstrate that it is possible to extrapolate accurately under multicollinearity with the least squares estimated model (\ref{ee}). Consider
\\[0.1in]
\noindent \hspace*{0.2in} $\mbox{\boldm $x$}_1=(7.46153, -11.76923, 48.15385, -30.00000),$ \\
\hspace*{0.2in}           $\mbox{\boldm $x$}_2=(3.18232, -15.98495, 64.86423, -10.86569),$\\
\hspace*{0.2in}           $\mbox{\boldm $x$}_3=(7.25776, -11.10359, 46.53671, -28.84034),$\\
\hspace*{0.2in}           $\mbox{\boldm $x$}_4=(-4.76478, -25.08204, 75.10608,  -1.00862),$\\
\hspace*{0.2in}           $\mbox{\boldm $x$}_5=(13.57470, -18.42563, 75.10608, -47.39482).$

\vspace{0.1in}

\noindent The standardized values of the 5 points are: \\[0.1in]
\noindent \hspace*{0.2in} $\mbox{\boldm $x$}_1'=(0.00,  0.00,  0.00,  0.00),$ \\
\hspace*{0.2in}           $\mbox{\boldm $x$}_2'=(-0.21, -0.19,  0.31,  0.33),$\\
\hspace*{0.2in}           $\mbox{\boldm $x$}_3'=(-0.01,  0.03, -0.03,  0.02),$\\
\hspace*{0.2in}           $\mbox{\boldm $x$}_4'=(-0.60, -0.60,  0.50,  0.50),$\\
\hspace*{0.2in}           $\mbox{\boldm $x$}_5'=(0.30, -0.30,  0.50, -0.30).$

\renewcommand{\baselinestretch}{1.0}
\begin{table}
\caption{\label{h3} Predicted values and their estimated variances at 5 points}
\centering
\fbox{ %
\begin{tabular}{ccc}
& Predicted value & Estimated variance\\
$\mbox{\boldm $x$}_1$ & 95.423   &  0.460\\
$\mbox{\boldm $x$}_2$ & 100.496 &  3.706\\
$\mbox{\boldm $x$}_3$ & 94.381  & 7.359\\
$\mbox{\boldm $x$}_4$ & 95.742  &  5.285\\
$\mbox{\boldm $x$}_5$ &  116.827 & 1689.129\\
\end{tabular}}
\end{table}
\renewcommand{\baselinestretch}{1.6}

\renewcommand{\baselinestretch}{1.0}
\begin{figure}
     \centering
     \includegraphics[scale=0.5]{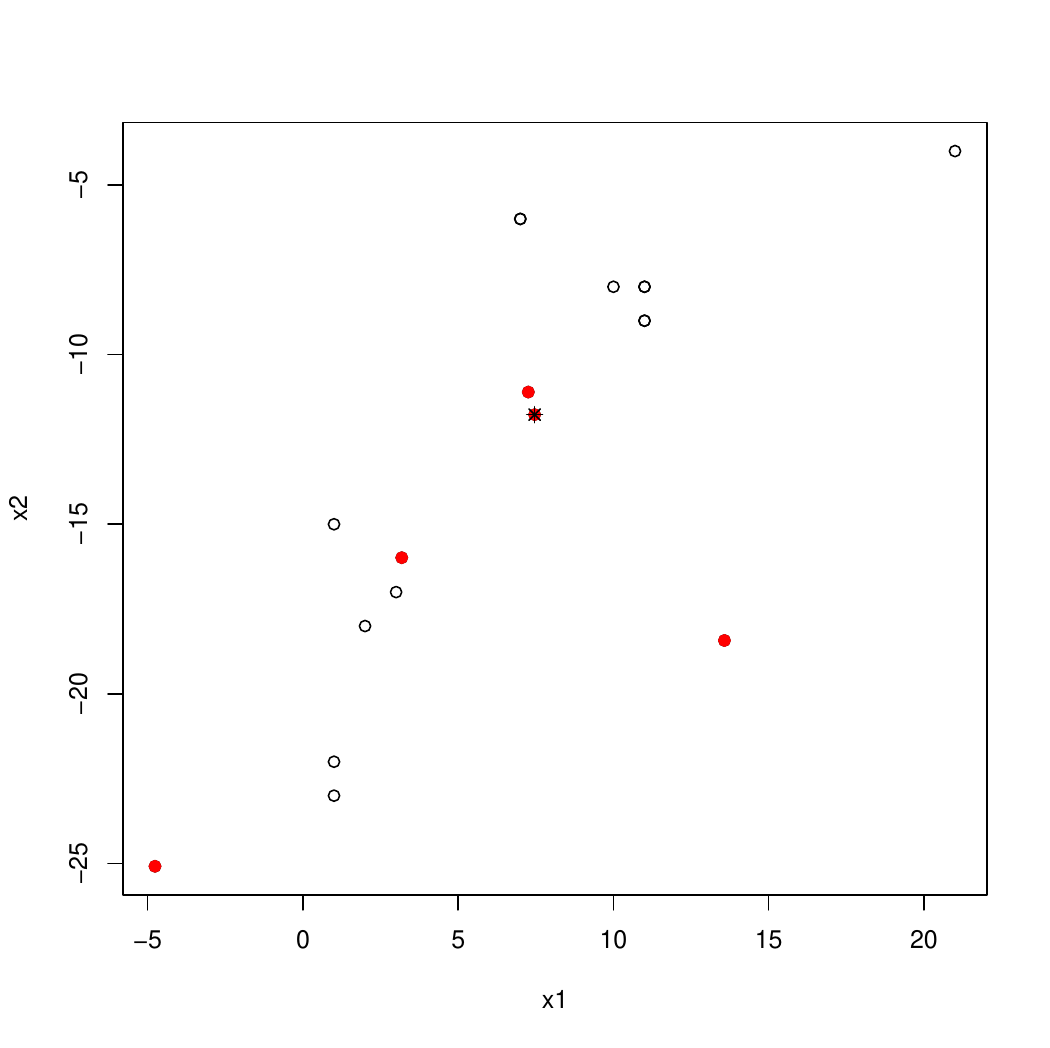}
     \caption{Points representing ($x_1,x_2)$ of the 13 observations in the Hald cement data are in circles. The ``$\star$'' symbol represents the mean of the 13 points. Points representing the 5 prediction points are in red dots. Points $\mbox{\boldm $x$}_4$ and $\mbox{\boldm $x$}_5$ are the two red dots outside the circle data hull, and $\mbox{\boldm $x$}_4$ is the one in the lower left corner which is still inside the feasible prediction region. A plot of ($x_3,x_4)$ of these points (not included) gives similar observations.}   \label{fig1}
\end{figure}
\renewcommand{\baselinestretch}{1.6}

\vspace{0.1in}

\noindent Since the strongly correlated groups in APC arrangement are $\{x_1,x_2\}$ and $\{x_3,x_4\}$, an $\mbox{\boldm $x$}_i$ is in ${\mathcal R}_{FP}$ if  its standardized version $\mbox{\boldm $x$}_i'=(x_1', x_2', x_3', x_4')$ satisfies $x_1'\approx x_2'$ and $x_3'\approx x_4'$. Thus, $\mbox{\boldm $x$}_1$, $\mbox{\boldm $x$}_2$, $\mbox{\boldm $x$}_3$ and $\mbox{\boldm $x$}_4$ are in ${\mathcal R}_{FP}$. Plotting  $(x_1,x_2)$ of the 5 points and the 13 points in the Hald cement data in Figure 1 finds $\mbox{\boldm $x$}_4$ and $\mbox{\boldm $x$}_5$ outside the data hull of the 13 points, so making predictions at $\mbox{\boldm $x$}_4$ and $\mbox{\boldm $x$}_5$ is extrapolation. Table \ref{h3} gives the predicted values and their estimated variances (\ref{evar}) at the 5 points. The predictions at $\mbox{\boldm $x$}_1$,  $\mbox{\boldm $x$}_2$ and  $\mbox{\boldm $x$}_3$ are accurate with small variances as these points are in both the data hull and ${\mathcal R}_{FP}$. Point $\mbox{\boldm $x$}_5$ is not in ${\mathcal R}_{FP}$ as it violated the strong positive correlation of the data (its $x_1'=0.3$ but $x_2'=-0.3$), so extrapolation at $\mbox{\boldm $x$}_5$ is highly inaccurate with a large variance. In contrast, extrapolation at $\mbox{\boldm $x$}_4$ is accurate as $\mbox{\boldm $x$}_4$ is in ${\mathcal R}_{FP}$. 
To summarize, extrapolation with the least squares estimated model can be accurate if it is done within the feasible prediction region.

%%%%%5555555555555555555555555555555555555555555555555555555555555555555555555555555555555

\section{Concluding remarks}

Multicollinearity due to strongly correlated predictor variables manifests in two ways. Numerically, it manifests through the ill-conditioning of the $\mathbf{X}^T\mathbf{X}$ matrix and ultimately the large variances of the least squares estimators for parameters of the strongly correlated variables. Geometrically, it manifests as a tight spacial constraint on the strongly correlated variables in that their data points are clustered tightly around a line\footnote{For unstandardised variables and/or variables not in an APC arrangement, this line is difficult to characterize. But for standardized variables in APC arrangement, this line is easy to describe; e.g., for the $q$ variable in $\mathbf{X}_1'$ of (\ref{m3}), this line is $x_1'=x_2'=\dots=x_q'$.}. Making predictions outside a narrow band around this line, including estimating parameters of these variables, is extreme extrapolation that may be meaningless and highly inaccurate.

Existing methods for dealing with multicollinearity such as ridge regression and principal component regression all focus on overcoming the numerical ill-conditioning aspect of multicollinearity in order to produce more accurate estimators for parameters of the strongly correlated variables. They overlook the geometric implication of multicollinearity which renders these parameters meaningless (see Remark $[a]$ of Section 2.3). They may produce estimators with smaller variances than the least squares estimators but this does not make the parameters they are trying to estimate more meaningful. Indeed, trying to accurately estimate parameters of strongly correlated variables is misguided. It also cannot be done in general as strongly correlated data contains little information about the individual parameters. With the misconception of their having more accurate predictions dispelled, there is little reason for abandoning the simple least squares regression in favour of these methods.

The group approach to the least squares regression respects the group nature of the strongly correlated predictor variables. It studies their group impact and is free of the multicollinearity problem. With the aid of the APC arrangement, it works effectively in estimation, inference, variable selection and prediction. We did not discuss model checking but on this point the group approach also has a clear advantage over the ridge regression and principal component regression as various residuals and residual plots for the least squares regression can be directly employed by the group approach with well-understood usages and interpretations, whereas the same cannot be said about the ridge regression and principal component regression. To conclude, we recommend the group approach to the least squares regression over existing methods for handling multicollinearity because of its simplicity and effectiveness.

%\appendix

\section{Appendix I: proofs of lemmas and theorems}\label{app}

\noindent {\bf Proof of Lemma \ref{lemma1}.}
Let $\mathbf{A}$ be the $q\times q$ matrix whose elements are all 1. Then, $\mathbf{A}$ has two distinct eigenvalues, $\lambda^A_1=q$ and $\lambda^A_2=0$. Eigenvalue $\lambda^A_1$ has multiplicity 1 and $\lambda^A_2$ has multiplicity $(q-1)$. The orthonormal eigenvector of $\lambda^A_1$ is $\frac{1}{\sqrt{q}}\mathbf{1}_q$. Here, we ignore the other orthonormal eigenvector of $\lambda^A_1$, $-\frac{1}{\sqrt{q}}\mathbf{1}_q$, which differs only in sign from $\frac{1}{\sqrt{q}}\mathbf{1}_q$.

Let $\mathbf{P}=[p_{ij}]$ be a perturbation matrix of $\mathbf{A}$ defined by
\beq
\mathbf{P}=\mathbf{A}-\mathbf{R}. \label{perb.matrix}
\eeq
Then, $\mathbf{P}$ is real and symmetric and $p_{ij}=1-r_{ij}$. When $r_M \rightarrow 1$, since $p_{ij}=(1-r_{ij})\rightarrow 0$, we have $\|\mathbf{P}\|_2\rightarrow 0$. It follows from this and $\mathbf{R}=\mathbf{A}-\mathbf{P}$ (so $\mathbf{R}$ is a perturbed version of $\mathbf{A}$) that $\lambda_1 \rightarrow \lambda_1^A=q$ and $\lambda_i \rightarrow \lambda^A_2=0$ for $i=2,3,\dots,q$  as $r_M \rightarrow 1$ (Horn and Johnson, 1985; page 367). 

To show that $\mathbf{v}_1\rightarrow \frac{1}{\sqrt{q}}\mathbf{1}_q$ as $r_M \rightarrow 1$, since $\mathbf{R} \mathbf{v}_1=\lambda_1\mathbf{v}_1$,  we have
\beq  r_{i1}v_{11}+r_{i2}v_{12}+\dots+r_{iq}v_{1q}=\lambda_1v_{1i} \label{t1} \eeq
for $i=1,2,\dots, q$, where $(r_{i1}, r_{i2}, \dots, r_{iq})$ is the $i$th row of $\mathbf{R}$ and $v_{1i}$ is the $i$th element of $\mathbf{v}_1$. All   $v_{1i}$ are bounded between $-1$ and $1$ since $v_{1i}^2 \leq  \|\mathbf{v}_1\|^2= 1$. When $r_M \rightarrow 1$,  all $r_{ij}\rightarrow 1$, so $(r_{ij}v_{1j}-v_{1j})\rightarrow 0$  for $j=1,2,\dots,q$. Thus,
%so the left-hand side of equation (\ref{t1}) satisfies
\beq
 (r_{i1}v_{11}+r_{i2}v_{12}+\dots+r_{iq}v_{1q})-(v_{11}+v_{12}+\dots+v_{1q}) \rightarrow 0   \label{t2}
\eeq
as $r_M \rightarrow 1$. By (\ref{t1}) and (\ref{t2}), $\lambda_1v_{1i} - (v_{11}+v_{12}+\dots+v_{1q}) \rightarrow 0$ which implies 
$\lambda_1^2v_{1i}^2 - (v_{11}+v_{12}+\dots+v_{1q})^2 \rightarrow 0$ for $i=1,2,\dots,q$. It follows that
\beq
 \lambda_1^2(v_{11}^2+v_{12}^2+\dots+v_{1q}^2)-q(v_{11}+v_{12}+\dots+v_{1q})^2 \rightarrow 0.   \label{t3}
\eeq
Since $v_{11}^2+v_{12}^2+\dots+v_{1q}^2= \|\mathbf{v}_1\|^2=1$ and $\lambda_1 \rightarrow q$, (\ref{t3}) implies that $(v_{11}+v_{12}+\dots+v_{1q})\rightarrow \sqrt{q}$. This and (\ref{t2}) imply that $$ (r_{i1}v_{11}+r_{i2}v_{12}+\dots+r_{iq}v_{1q}) \rightarrow \sqrt{q}$$
for $i=1,2,\dots,q$. By  (\ref{t1}), we also have $\lambda_1v_{1i} \rightarrow \sqrt{q}$. This and  $\lambda_1 \rightarrow q$ imply that $v_{1i}\rightarrow 1/\sqrt{q}$ for $i=1,2,\dots,q$, that is,
$\mathbf{v}_1\rightarrow \frac{1}{\sqrt{q}}\mathbf{1}_q$.    \hfill $\square$

\vspace{0.2in} %%%%%%%%%%%%%%%%%%%%%%%%%%%%%%%%%%%%%%%   Lemma 2

\noindent {\bf Proof of Lemma \ref{lemma2}.}
Since $\mathbf{R}$ is positive definite, $\mathbf{R}^{-1}$ is also positive definite. Let $\lambda_1'\geq \lambda_{2}'\geq \dots\geq \lambda_q'>0$ be the eigenvalues of $\mathbf{R}^{-1}$. Then, $\lambda_i'=\lambda^{-1}_{q-i+1}$ and its eigenvector is $\mathbf{v}_i'=\mathbf{v}_{q-i+1}$ for $i=1,2, \dots, q$. In particular, $\lambda_q'=\lambda_1^{-1}$ and $\mathbf{v}_q'=\mathbf{v}_1$. Since all $\lambda_i>0$ and $trace(\mathbf{R})=q=\sum^q_{i=1}\lambda_i$, we have $0<\lambda_1<q$. Also, $\mathbf{v}_1^T\mathbf{v}_1=1$ as $\mathbf{v}_1$ is orthonormal. It follows from these that
\begin{equation}
 \mathbf{v}_1^T \mathbf{R}^{-1} \mathbf{v}_1 =  \mathbf{v}_q'^T \mathbf{R}^{-1} \mathbf{v}_q' =  \mathbf{v}_q'^T\lambda_q' \mathbf{v}_q' =
\frac{\mathbf{v}_1^T\mathbf{v}_1 }{\lambda_1}=\frac{1}{\lambda_1} >\frac{1}{q},  \label{temp10}
 \end{equation}
which proves ($i$). By Lemma 1, $\lambda_1 \rightarrow q$ as $r_M \rightarrow 1$. Thus, by (\ref{temp10})
\[ \mathbf{v}_1^T \mathbf{R}^{-1} \mathbf{v}_1 = \frac{1}{\lambda_1} \rightarrow \frac{1}{q}, \]
as $r_M \rightarrow 1$, which proves ($ii$). \hfill $\square$

\hspace{0.1in}  %%%%%%%%%%%%%%%%%%%%%%%%%%%%%%%%%%%%%%%%%   Theorem 1

\noindent {\bf Proof of Theorem \ref{thm1}.}
For any constant vector $\mathbf{c} \in \mathbb{R}^p$, we have
\beq
{var}(\mathbf{c}^T\hat{\boldsymbol{\beta}}')={\sigma}^2\mathbf{c}^T[\mathbf{X}'^T\mathbf{X}']^{-1}\mathbf{c}. \label{var2}
\eeq
Let $\mathbf{c}_E=({\mathbf{v}^*_1}^T, 0, \dots, 0)^T$. Then, ${\xi}_E= \mathbf{c}_E^T {\boldsymbol{\beta}}'$ and $\hat{\xi}_E= \mathbf{c}_E^T \hat{\boldsymbol{\beta}}'$. 
By (\ref{corr.inv}) and (\ref{var2}),
\beq
var(\hat{\xi}_E)
=\sigma^2{\mathbf{v}^*_1}^T[\mathbf{R}_{11}-\mathbf{R}_{12}\mathbf{R}^{-1}_{22}\mathbf{R}_{21}]^{-1}{\mathbf{v}^*_1}
= \sigma^2 {\mathbf{v}^*_1}^T{\mathbf{R}^*}^{-1}{\mathbf{v}^*_1}. \label{var3}
\eeq 
To show ($i$), when variables in $\mathbf{X}_1'$ are uncorrelated with variables in $\mathbf{X}_2'$, $\mathbf{R}_{12}=\mathbf{0}$ and so $\mathbf{R}^*=\mathbf{R}$ and $\mathbf{v}^*_1=\mathbf{v}_1$.   By (\ref{var3}),
\beq
var(\hat{\xi}_E)
={\sigma}^2\mathbf{v}^T_1 \mathbf{R}^{-1}\mathbf{v}_1. \label{var4}
\eeq
Applying Lemma 2 to the right-hand side of (\ref{var4}), we obtain ($i_1$) and ($i_2$).

To show ($ii$),  for simplicity we assume general conditions discussed in footnote 1 hold so that $\mathbf{R}_{12}\mathbf{R}^{-1}_{22}\mathbf{R}_{21}\rightarrow \mathbf{0}$ when $\mathbf{R}_{12}\rightarrow \mathbf{0}$.  It follows from this and conditions in Theorem \ref{thm1}($ii$) that $\mathbf{R}_{11}$ and $\mathbf{R}^*$ will both converge to matrix $\mathbf{A}$  in (\ref{perb.matrix}). We again define a perturbation matrix of $\mathbf{A}$ as
$$ \mathbf{P}^*=\mathbf{A} - \mathbf{R}^* $$
like what we did in (\ref{perb.matrix}). By following steps similar to those in the proofs of Lemma \ref{lemma1} and Lemma \ref{lemma2}, we can show that $\mathbf{R}^*$ also has the two properties  in Lemma \ref{lemma1} and property ($ii$) in Lemma \ref{lemma2}. The latter and (\ref{var3}) imply ($ii$).  \hfill $\square$

\hspace{0.1in}  %%%%%%%%%%%%%%%%%%%%%%%%%%%%%%%%%%%%%%%%%   Theorem 2

\noindent {\bf Proof of Theorem \ref{thm2}.}  Since $\mathbf{v}\cdot \mathbf{v}^*_1=\|\mathbf{v}\|\|\mathbf{v}^*_1\|cos(\theta)=cos(\theta)$ where $\theta$ is the angle between $\mathbf{v}$ and $\mathbf{v}^*_1$, $\sqrt{1-\delta}< \mathbf{v} \cdot \mathbf{v}^*_1 \leq 1$ is equivalent to  $\sqrt{1-\delta}<cos(\theta) \leq 1$ or $0\leq \theta< \theta_\delta$ for some small fixed $\theta_\delta>0$. Thus, ${\cal N}_{\delta}$ in (\ref{temp11}) represents a small open circular region centred on $\mathbf{v}_1^*$ on the surface of the unit sphere. 

Similar to $var(\hat{\xi}_E)$ in (\ref{var3}), $var(\mathbf{v}^T\hat{\boldsymbol{\beta}}_1')=\sigma^2\mathbf{v}^T{\mathbf{R}^*}^{-1}\mathbf{v}$. Since  ${\mathbf{R}^*}^{-1}$ is real symmetric positive definite, it has eigendecomposition $\mathbf{Q\Lambda Q}^T$ where $\mathbf{Q}$ is the matrix of orthonormal eigenvectors including $\mathbf{v}^*_1$ and $\mathbf{\Lambda}$ is the diagonal matrix of eigenvalues. The smallest eigenvalue of ${\mathbf{R}^*}^{-1}$ is  $1/\lambda^*_1$ which converges to $1/q$ under the condition of Theorem \ref{thm2} as $r_M$ goes to 1. The other eigenvalues of ${\mathbf{R}^*}^{-1}$ all go to infinity  as $r_M$ goes to 1. For any unit vector $\mathbf{v}$, 
\begin{equation}
1=\mathbf{v}^T\mathbf{v} =  \mathbf{v}^T\mathbf{Q}\mathbf{Q}^T\mathbf{v}= 
 \mathbf{v}^T[\tilde{\mathbf{Q}},\mathbf{v}^*_1][\tilde{\mathbf{Q}},\mathbf{v}^*_1]^T\mathbf{v} 
= \mathbf{v}^T\tilde{\mathbf{Q}}\tilde{\mathbf{Q}}^T\mathbf{v}+ (\mathbf{v}^T \mathbf{v}^*_1)^2      \label{32}
\end{equation}
where $\tilde{\mathbf{Q}}$ is the matrix containing all columns of $\mathbf{Q}$ but $\mathbf{v}^*_1$. If $\mathbf{v} \notin {\mathcal N}_{\delta}$, then $(\mathbf{v}^T \mathbf{v}^*_1)^2\leq 1-\delta$. This and (\ref{32}) imply that $1\leq \mathbf{v}^T\tilde{\mathbf{Q}}\tilde{\mathbf{Q}}^T\mathbf{v}+(1-\delta)$, that is, $ \mathbf{v}^T\tilde{\mathbf{Q}}\tilde{\mathbf{Q}}^T\mathbf{v} \geq \delta$. This leads to the following lower bound on $var(\mathbf{v}^T\hat{\boldsymbol{\beta}}_1')$,
\begin{equation}
 var(\mathbf{v}^T\hat{\boldsymbol{\beta}}_1')=\sigma^2\mathbf{v}^T{\mathbf{R}^*}^{-1}\mathbf{v}
= \sigma^2\mathbf{v}^T\mathbf{Q\Lambda Q}^T\mathbf{v}\geq \sigma^2\mathbf{v}^T\tilde{\mathbf{Q}} \tilde{\mathbf{\Lambda}} \tilde{\mathbf{Q}}^T\mathbf{v}\geq \frac{\sigma^2\delta}{\lambda^*_2},     \label{33}
\end{equation}
where $ \tilde{\mathbf{\Lambda}}$ is the diagonal matrix of all eigenvalues of  ${\mathbf{R}^*}^{-1}$ except the smallest one $1/\lambda^*_1$, and $1/\lambda^*_2$ is the second smallest eigenvalue of ${\mathbf{R}^*}^{-1}$. Since $1/\lambda^*_2 \rightarrow \infty$ as $r_M \rightarrow 1$, (\ref{33}) implies that $var(\mathbf{v}^T\hat{\boldsymbol{\beta}}_1') \rightarrow \infty$ as $r_M \rightarrow 1$ if $\mathbf{v} \notin {\mathcal N}_{\delta}$. 
 \hfill $\square$
 
 \section{Appendix II: multicollinearity in generalized linear models}\label{app}
 
Multicollinearity can also affect estimation and inference for generalized linear models. In this section, we examine its impact through a non-rigorous analysis of an example involving a logistic regression model. We again focus on the variance of the estimator for group effects of strongly correlated predictor variables.
 
Let $Y_1, Y_2, \dots, Y_n$ be $n$ independent observations of the response variable where $Y_i \sim \mbox{Binomial}(m, \pi_i)$, and let $\mathbf{X}=[ \mathbf{x}_1,\dots,\mathbf{x}_{p}]$ be the corresponding $n\times p$ matrix of predictor variables where $p>2$. We assume that ($i$) all variables $\mathbf{x}_i$ are standardized variables with mean zero and length one, ($ii$) $\{\mathbf{x}_{1}, \mathbf{x}_{2}\}$ is a group of strongly correlated variables in an APC arrangement, and ($iii$) $\mathbf{x}_{1}$ and $\mathbf{x}_{2}$ are weakly correlated with $\mathbf{x}_i$ for $i \in \{3, 4, \dots, p\}$.
The logistic regression model is given by
\beq
	logit(\pi_i)=\mathbf{x}^r_{i}\boldsymbol{\beta},  \label{logit}
\eeq
or alternatively,
\beq
	\pi_i=\frac{\exp(\mathbf{x}^r_{i}\boldsymbol{\beta})}{1+ \exp(\mathbf{x}^r_{i}\boldsymbol{\beta})} , \nonumber  %\label{logistic}
\eeq
where $\mathbf{x}^r_{i}$ is the $i$th row of the design matrix $\mathbf{X}$. Let $\hat{\boldsymbol{\beta}}$ be the maximum likelihood estimator of $\boldsymbol{\beta}$. The asymptotic variance matrix of $\hat{\boldsymbol{\beta}}$ is
\beq
Var( \hat{\boldsymbol{\beta}}) = (\mathbf{X}^T\mathbf{V}\mathbf{X})^{-1},    \label{asy.var}
\eeq
where $\mathbf{V}=diag(v_1, v_2, \dots, v_n)$ is a diagonal matrix with $v_i=Var(Y_i)$.  Let
\[
\mathbf{U}=\mathbf{X}^T\mathbf{V}\mathbf{X}=[{u}_{ij}]_{p \times p}.
\]
Then,
\beq
{u}_{ij}= \mathbf{x}_i^T \mathbf{V}  \mathbf{x}_j= \sum^n_{k=1} (x_{ik}x_{jk})v_k. \label{ele1}
\eeq
When $i=j$, we have
 \beq
{u}_{ii}= \mathbf{x}_i^T \mathbf{V}  \mathbf{x}_i= \sum^n_{k=1} (x_{ik}x_{ik})v_k=\sum^n_{k=1} x_{ik}^2v_k.   \label{ele2}
\eeq
Suppose the distributions of $\mathbf{x}_i$ are the same for all $i$, then $u_{ii}$ are roughly the same for all $i$ and they should be large when $m$ and $n$ are large.
To assess the relative size between $u_{ii}$ and $u_{ij}$ where $i\neq j$, we consider the following two cases.  

Case (I): $i=1$ and $j=2$. Since $\mathbf{x}_1$ and $\mathbf{x}_2$ are standardized variables that are strongly and positively correlated, they are approximately equal; that is, $x_{1k} \approx x_{2k}$ for $k=1,2,\dots, n$. This implies $x_{1k} x_{2k}\approx x_{1k}^2$, so by (\ref{ele1}) and (\ref{ele2}), ${u}_{12}\approx {u}_{11}$. Similarly, we have $u_{11}\approx {u}_{21}=u_{12} \approx {u}_{22}$. Partition the $\mathbf{U}$ matrix as follows,
\beq
\mathbf{U}=
\left[ \begin{array}{cc}
\mathbf{U}_{11} &\mathbf{U}_{12} \\
\mathbf{U}_{21} & \mathbf{U}_{22}  \\
\end{array} \right]_{(p+1)\times (p+1)},  \label{xvx} 
\eeq
where 
\beq
\mathbf{U}_{11}=
\left[ \begin{array}{cc}
{u}_{11} & {u}_{12} \\
u_{21} & {u}_{22}  \\
\end{array} \right]_{2\times 2}.  \label{xvx2} 
\eeq
Then, elements in $\mathbf{U}_{11}$ are all approximately equal to $u_{11}$. 

Case (II): $i=1$ and $j \in \{3, 4, \dots, p\}$. Since $\mathbf{x}_1$ and $\mathbf{x}_j$ are weakly correlated standardized variables, $\sum^n_{k=1}x_{1k} x_{jk} \approx 0$. This and the observation that $v_k$ are largely independent of $(x_{1k}x_{jk})$ suggest that $\sum^n_{k=1} (x_{1k}x_{jk})v_k \approx  0$. Thus,
\beq
{u}_{11}= \sum_{k=1}^n x_{1k}^2v_k \gg u_{1j}= \sum^n_{k=1} (x_{1k}x_{jk})v_k \approx  0.    \label{tt2}
\eeq
Similarly, ${u}_{j1}, {u}_{2j}$ and ${u}_{j2}$ are also expected to be small in terms of absolute value relative to $u_{11}$. To summarize,
the above analysis shows that elements of $\mathbf{U}_{11}$ are roughly equal, and by (\ref{tt2}), they are large relative to elements in $\mathbf{U}_{12}=\mathbf{U}_{21}^T$.

By (\ref{xvx}), the inverse of $\mathbf{U}$ is
\beq
\mathbf{U}^{-1}=
\left[ \begin{array}{cc}
[\mathbf{U}_{11}-\mathbf{U}_{12}\mathbf{U}^{-1}_{22}\mathbf{U}_{21}]^{-1} &  \mathbf{U}_{11}^{-1}\mathbf{U}_{12}[\mathbf{U}_{21}\mathbf{U}^{-1}_{11}\mathbf{U}_{12} - \mathbf{U}_{22}]^{-1}  \\

[\mathbf{U}_{21}\mathbf{U}^{-1}_{11}\mathbf{U}_{12} - \mathbf{U}_{22}]^{-1} \mathbf{U}_{21} \mathbf{U}_{11}^{-1} & [\mathbf{U}_{22}-\mathbf{U}_{21}\mathbf{U}^{-1}_{11}\mathbf{U}_{12}]^{-1} \\
\end{array} \right].  \label{xvx.inv}
\eeq
Let $\xi(\mathbf{w})=w_1\beta_1+w_2\beta_2$ be a group effect of $\{\mathbf{x}_1, \mathbf{x}_2\}$ where $\mathbf{w}=(w_1, w_2)^T$. Then, the asymptotic variance of its estimator
$\hat{\xi}(\mathbf{w})=w_1\hat{\beta}_1+w_2\hat{\beta}_2$ is
\[
Var(\hat{\xi}(\mathbf{w}))= \mathbf{w}^T[\mathbf{U}_{11}-\mathbf{U}_{12}\mathbf{U}^{-1}_{22}\mathbf{U}_{21}]^{-1} \mathbf{w}.
\]
Since elements in  $\mathbf{U}_{12}=\mathbf{U}_{21}^T$ are small, elements of $\mathbf{U}_{12}\mathbf{U}^{-1}_{22}\mathbf{U}_{21}$ should also be small. This and the observation that elements of $\mathbf{U}_{11}$ are relatively large and approximately equal to $u_{11}$ imply that 
\[
\mathbf{U}^*_{11}=u_{11}^{-1}[\mathbf{U}_{11}-\mathbf{U}_{12}\mathbf{U}^{-1}_{22}\mathbf{U}_{21}] \approx [1]_{2\times 2},
\]
that is, $\mathbf{U}^*_{11}$ is a perturbed version of the $\mathbf{A}$ matrix in (\ref{perb.matrix}) for $q=2$. Thus, Lemma 1 and Lemma 2($ii$) may be applicable to $\mathbf{U}^*_{11}$.
Noting that
\[
Var(\hat{\xi}(\mathbf{w}))= \mathbf{w}^T[\mathbf{U}_{11}-\mathbf{U}_{12}\mathbf{U}^{-1}_{22}\mathbf{U}_{21}]^{-1} \mathbf{w}=\frac{1}{u_{11}} \mathbf{w}^T[\mathbf{U}^*_{11}]^{-1} \mathbf{w}
\]
which is similar to (\ref{var3}), we see that similar to the linear model case, the normalized eigen-effect of $\mathbf{U}_{11}^*$ and thus the average group effect $\xi_A=\frac{1}{2}(\beta_1+\beta_2)$ may be accurately estimated. Other effects in the neighbourhood of $\xi_A$ may also be accurately estimated, but $\beta_1$ and $\beta_2$ will not be as $Var(\hat{\beta}_1)$ and $Var(\hat{\beta}_2)$ are large.

In the above example, we considered only 2 strongly correlated predictor variables but the same steps may be applied to a group of $q$ such variables to obtain similar observations.
This example revealed two difficulities in studying the impact of multicollinearity on generalized linear models: ($i$) the finite sample variance formula for $\hat{\boldsymbol{\beta}}$ is not available, so we have to rely on the asymptotic variance formula $(\mathbf{X}^T\mathbf{V}\mathbf{X})^{-1}$, and ($ii$) the asymptotic variance formula depends on $\mathbf{V}$ which varies from one generalized linear model to another. Because of problem ($i$), we have no means to make precise assessment of the impact of multicollinearity on generalized linear models for finite sample situations. Assessments obtained through analyzing the asymptotic variance formula are likely to under-estimate the real impact in small and moderate sample situations as the asymptotic variance formula is for large sample situations where multicollinearity is often not a serious problem. Problem ($ii$) makes it more difficult to study the limiting behaviour of the variance of the estimated group effects; e.g., in (\ref{ele2}), $u_{ii}$ would have been exactly 1 for all $i$ without the $\mathbf{V}$ matrix (or equivalently if $\mathbf{V}=\mathbf{I}$ as in the linear models), and in (\ref{tt2}), the comparison between $u_{11}$ and $u_{1j}$ would have been easier without the $\mathbf{V}$ matrix. On the other hand, for cases where the asymptotic variance formula is valid, multicollinearity usually causes less trouble for generalized linear models when compared to linear models; that is, given the same level of multicollinearity in the design matrix $\mathbf{X}$, matrix $ (\mathbf{X}^T\mathbf{V}\mathbf{X})$ is usually not as ill-conditioned as matrix $(\mathbf{X}^T\mathbf{X})$; the addition of the $\mathbf{V}$ matrix in the variance formula has reduced the impact of multicollinearity.


\begin{thebibliography}{9}


\bibitem{Belsley2004}
Belsley, D. A., Kuh, E., Welsch, R. E. (2004).
{\em Regression Diagnostics: Identifying Influential Data and Sources of Collinearity},
Wiley \& Sons, New York

\bibitem{Conniffe1973}
Conniffe, D., Stone, J. (1973).
{A critical view of ridge regression}.
{\em American Statistician}, {22}, 181--187

\bibitem{Dampster1977}
Dampster, A. P., Schatzoff, M., Wermuth, N. (1977).
{A simulation study of alternatives to ordinary least squares},
{\em Journal of the American Statistical Association}, {72}, 77--90

\bibitem{Draper1998}
Draper, N. R., Smith, H. (1998).
{\em Applied Regression Analysis}, 3rd ed.,
Wiley, New York

\bibitem{Draper1979}
Draper, N. R., Van Nostrand, R. C. (1979).
{Ridge regression and James-Stein estimators: Review and comments},
{\em Technometrics}, {21} 451--466

\bibitem{Friedman2017}
Friedman, J., Hastie, T., Simon, N., Tibshirani, R. (2017).
{Package `glmnet'}, an R package available at
https://cran.r-project.org

\bibitem{Gunst1977}
Gunst, R. F., Mason, R. L. (1977).
{Biased estimation in regression: An evaluation using mean squared error}.
{\em Journal of the American Statistical Association}, {72}, 616--628

\bibitem{Gunst1976}
Gunst, R. F., Webster, J. T., Mason, R. L. (1976).
{A comparison of least squares and latent root regression estimators},
{\em Technometrics}, {18}, 75--83

\bibitem{Hoerl1970}
Hoerl, A. E., Kennard, R. W. (1970).
{Ridge Regression: biased estimation for nonorthonogal problems},
{\em Technometrics}, {12}, 55--67

\bibitem{Hoerl1975}
Hoerl, A. E., Kennard, R. W., Baldwin, K. F. (1975).
{Ridge regression: Some simulations},
{\em Communications in Statistics: Theory and Methods}, {4}, 105--123

\bibitem{Horn1985}
Horn, R. A., Johnson, C. A. (1985).
{\em Matrix Analysis},
Cambridge University Press
%\MR{838085}

\bibitem{Jolliffe1986}
Jolliffe, I. T. (1986).
{\em Principal component analysis.} 
Springer-Verlag, New York

\bibitem{Lawless1978}
Lawless,  J. F. (1978).
{Ridge and related estimation procedures: Theory and practice},
{\em Communications in Statistics: Theory and Methods}, {7}, 135--164

\bibitem{Lawless1976}
Lawless, J. F., Wang, P. (1976).
{A simulation study of ridge and other regression estimators},
{\em Communications in Statistics: Theory and Methods}, {5}, 307--323

\bibitem{Lumley2017}
Lumley, T., Miller, A. (2017).
{Package `leaps'}, an R package available at
https://cran.r-project.org

\bibitem{Montgomery2012}
Montgomery, D. C., Peck, E. A., Vining, G. G. (2012).
{\em Introduction to Linear Regression Analysis}, 5th ed.,
Wiley, New York

\bibitem{Tsao2019}
Tsao, M. (2019).
{Estimable group effects for strongly correlated variables in linear models},
{\em Journal of Statistical Inference and Planning}, {198}, 29--42

\bibitem{temp}
Webster J. T., Gunst, R. F., Mason, R. L. (1974).
{Latent root regression analysis},
{\em Technometrics}, {16}, 513--522


\end{thebibliography}
\end{document}